\documentclass{emulateapj}
\usepackage{natbib}
\slugcomment{To appear in the Astrophysical Journal}
\shortauthors{Fitzpatrick \& Massa}
\shorttitle{IR Extinction}

\begin{document}

\defcitealias{FMI}{Paper~I}
\defcitealias{FMII}{Paper~II}
\defcitealias{FMIII}{Paper~III}
\defcitealias{FMIV}{Paper~IV}
\defcitealias{FMV}{Paper~V}
\defcitealias{mathis1981}{MW81}
\defcitealias{MRN}{MRN}

\newcommand{\atlas}{{ATLAS9}}
\newcommand{\tlusty}{{TLUSTY}}
\newcommand{\synspec}{{SYNSPEC}}
\newcommand{\iras}{{\it IRAS}}
\newcommand{\ans}{{\it ANS}}
\newcommand{\hst}{{\it HST}}
\newcommand{\iue}{{\it IUE}}
\newcommand{\oao}{{\it OAO-2}}
\newcommand{\td}{{\it TD-1}}
\newcommand{\tmass}{{2MASS}} 
\newcommand{\jhk}{{\it JHK}}  
\newcommand{\ubv}{{\it UBV}}                                                                                                                                                                                                                            
\newcommand{\mast}{{MAST}}
\newcommand{\irsa}{{IRSA}}
\newcommand{\hip}{{\it Hipparcos}}
\newcommand{\simbad}{{\it SIMBAD}}
\newcommand{\teff}{\mbox{$T_{\rm eff}$}}
\newcommand{\logg}{{$\log g$}}
\newcommand{\vturb}{$v_{turb}$}
\newcommand{\abund}{[m/H]}
\newcommand{\vsini}{$v \sin i$}
\newcommand{\kms}{km\,s$^{-1}$}
\newcommand{\msun}{${\rm M}_\sun$}
\newcommand{\ebv}{\mbox{$E(B\!-\!V)$}}
\newcommand{\klam}{\mbox{$k(\lambda \!-\!V)$}}
\newcommand{\mic}{\mbox{$\mu{\rm m}$}}
\newcommand{\invmic}{\mbox{$\mu{\rm m}^{-1}$}}

\title{An Analysis of the Shapes of Interstellar Extinction Curves. VI.
The Near-IR Extinction Law}

\author{E.L.~Fitzpatrick\altaffilmark{1}, D.~Massa\altaffilmark{2}}
\altaffiltext{1}{Department of Astronomy \& Astrophysics, Villanova
University, 800 Lancaster Avenue, Villanova, PA 19085, USA; 
edward.fitzpatrick@villanova.edu}
\altaffiltext{2}{SGT, Inc., Space Telescope Science Institute, 3700 San 
Martin Dr., Baltimore, MD 21218; massa@derckmassa.net}

\begin{abstract}
We combine new HST/ACS observations and existing data to investigate the wavelength dependence of near-IR (NIR) extinction.  Previous studies suggest a power-law form for NIR extinction, with a ``universal'' value of the exponent, although some recent observations indicate that significant sight line-to-sight line variability may exist.  We show that a power-law model for the NIR extinction provides an excellent fit to most extinction curves, but that the value of the power, $\beta$, varies significantly from sight line-to-sight line.  Therefore, it seems that a ``universal NIR extinction law'' is not possible.  Instead, we find that as $\beta$\ decreases, $R(V) \equiv A(V)/E(B-V)$ tends to increase, suggesting that NIR extinction curves which have been considered ``peculiar'' may, in fact, be typical for different $R(V)$\ values.

We show that the power law parameters can depend the wavelength interval used to derive them, with the $\beta$\ increasing as longer wavelengths are included.  This result implies that extrapolating power law fits to determine $R(V)$ is unreliable.  To avoid this problem, we adopt a different functional form for NIR extinction.  This new form mimics a power law whose exponent increases with wavelength, has only 2 free parameters, can fit all of our curves over a longer wavelength baseline and to higher precision, and produces $R(V)$\ values which are consistent with independent estimates and commonly used methods for estimating $R(V)$.  Furthermore, unlike the power law model, it gives $R(V)$s that are independent of the wavelength interval used to derive them.  It also suggests that the relation $R(V) = -1.36 \frac{E(K-V)}{E(B-V)} - 0.79$ can estimate $R(V)$ to $\pm 0.12$.

Finally, we use model extinction curves to show that our extinction curves are in accord with theoretical expectations, and demonstrate how large samples of observational quantities can provide useful constraints on the grain properties.
\end{abstract}


\section{INTRODUCTION\label{sec_INTRO}}
In the previous papers in this series, we began with an in-depth look at
the structure and properties of UV extinction curves, including the 2175
\AA\/ ``bump'' \citep{FMI, FMII, FMIII}, and most recently completed a
general survey of Galactic extinction covering the near IR (NIR) through 
UV spectral regions \citep[hereafter Papers IV and V, respectively]{FMIV,
FMV}. These studies have utilized \tmass\/ \jhk\/ photometry, optical
photometry, and low-resolution {\it International Ultraviolet Explorer}
satellite (\iue) spectrophotometry. Our general goal of codifying the
behavior of interstellar extinction over the widest possible wavelength
range has been motivated by two scientific objectives: (1) to aid in the
identification of the dust grain populations which produce the extinction
and (2) to provide tools for the removal of the effects of
wavelength-dependent extinction from astronomical data.

During the course of this study, it has become clear that our ability to
characterize extinction properties is strongly limited by the non-uniform
type and quality of the data available in the various spectral domains. In
fact, due to the success of the \iue\/ mission, the best characterized
extinction curves are in the once-inaccessible UV region (1200--3000 \AA),
while the worst-characterized regions are the ``gaps'' between UV and
optical data ($\sim$3000-3800 \AA) and between IR and optical data
($\sim$6000--10000 \AA). These regions are, in principle, accessible from
the ground but, due to calibration issues, remain poorly studied. In an
attempt to remedy this situation, we obtained spectrophotometric data from
the Advanced Camera for Surveys (ACS) aboard the {\it Hubble Space
Telescope} (\hst). The observations included both gap regions although, for
reasons discussed in \S~\ref{sec_ACS} below, only the data spanning the
optical-IR gap were useful for the program. These data, however, turn out 
to be of value not only for characterizing the wavelength-dependence of
extinction along the chosen set of sight lines, but also for addressing the
broader issue of a ``universal'' extinction law in the IR region. A variety
of observations have suggested a common form to extinction at wavelengths
from the $I$ band through $\backsim$5 $\mu$m \citep[e.g.,][]{rieke1985,
martin1990}, consistent with a power-law form for the extinction law,
$A_\lambda \propto \lambda^{-\beta}$, and an exponent value of $\beta =
1.84$ for both diffuse and dark cloud lines of sight \citep{whittet2003}.
More recent observations, however, have suggested significant
sight line-to-sight line variability in IR extinction with a potentially
large range of power-law exponents \citep{larson2005, nishiyama2006,
froebrich2007, gosling2009}.

In this paper we use our new ACS observations in the gap region between the
optical and the NIR, along with existing data, to investigate the shape of
the extinction law in this region and also the issue of whether a universal
NIR law exists.  In \S~2 the target stars and the data used
here (both new and archival) are described.  In \S~3, we 
discuss the technique used to derive extinction curves, and consider two 
families of analytic functions that can be used to represent them.  In 
\S~4, we examine how our results compare to theoretical 
models.  \S~5 summarizes the main results of the paper, 
and an appendix addresses technical issues concerning the ACS calibration. 

\section{THE TARGET STARS AND THEIR DATA\label{sec_TARGETS}}

In this study, we examine extinction curves towards 14 stars drawn from
a set of more than 300 stars studied in Paper V. These 14 stars, along with their
spectral types, $V$ magnitudes, and reddenings, are listed in Table 1. Since our method for deriving extinction
curves involves modeling the observed spectral energy distributions (SEDs)
of reddened stars --- using stellar atmosphere calculations and an analytical
form of the extinction curve --- we require absolutely-calibrated photometry
and spectrophotometry. The observing program and the processing of the new
\hst/ACS data are described in \S~\ref{sec_ACS}. The other data sets
utilized here \citepalias[via the analysis in][]{FMV} are described in \S~\ref{sec_OTHERDATA} and the complete energy distributions are presented in
\S~\ref{sec_ALLTOGETHER}.

\subsection{New ACS Observations\label{sec_ACS}}

The new observations were part of our Cycle 14 \hst/ACS program entitled 
``A SNAP Program to Obtain Complete Wavelength Coverage on Interstellar 
Extinction.'' Our goal was to obtain ACS/HRC spectra in the near-UV (using 
the PR200L grating) and in the NIR (using the GR800L).  These data, 
when combined with existing NIR ({\it JHK}) photometry, optical 
photometry, and \iue\/ UV 
spectrophotometry, would provide complete extinction curves over the range 
$\sim$1150 \AA\/ to $\sim$2.2 $\mu$m.  We submitted a target list of 50 
stars drawn from the sample in Paper V.  Ultimately, 22 stars were observed 
during 2005 and 2006 and, of these, 14 were deemed suitable for this study.  
The ACS data were processed and calibrated using the aXe Spectral 
Extraction package (version 1.6), developed by M. K\"{u}mmel, J. Walsh, 
and H. Kuntschner.  We obtained the package from the {\it Hubble Space 
Telescope} European Homepage\footnotemark 
\footnotetext{$www.stecf.org/software/slitless$\_$software/axe/$}.  The aXe 
software was run in conjunction with the IRAF/PYRAF package on a Sun 
Microsystems SunBlade1000 machine.  

Obtaining absolutely-calibrated spectrophotometry with the ACS is a
challenging process in general and for our HRC/PR200L observations it
proved to be impossible. The problem was the ``red pile-up'' effect whereby
all photons between 4000 and 10000 \AA\/ are focused onto a region of the
detector only 7 pixels wide \citep{larsen2006}. This is particularly
troublesome for red objects, such as our reddened targets, since
diffraction spikes from the resultant bright spot overlap the blue region
of the spectra. We found it impossible to extract reliably-uncontaminated
spectra from our HRC/PR200L observations and ultimately excluded these data
-- and consideration of the UV-visual extinction gap -- from this study.

Processing the HRC/G800L observations was more straightforward and we
followed the procedures outlined in the aXe Users Manual using an
extraction slit height of $\pm$20 pixels. The result was a set of
absolutely-calibrated first-order grism spectra for our 14 targets covering
the range 5700-10200 \AA. Because absolute calibration is critical to our
program, we also verified the HRC/G800L calibration by examining the
HRC/G800L spectra of three DA white dwarf stars used to establish the
calibration. We found a significant systematic discrepancy between the
calibration targets processed by aXe and their corresponding model
atmospheres. Consequently, we had to derive a correction factor for our
spectra. This calibration correction is provided in the Appendix. The final
step in our ACS data processing was to trim the spectra to the wavelength
range 6000--9500 \AA, where the fluxes are most reliable.

\begin{figure}[ht]
\figurenum{1}
\plotone{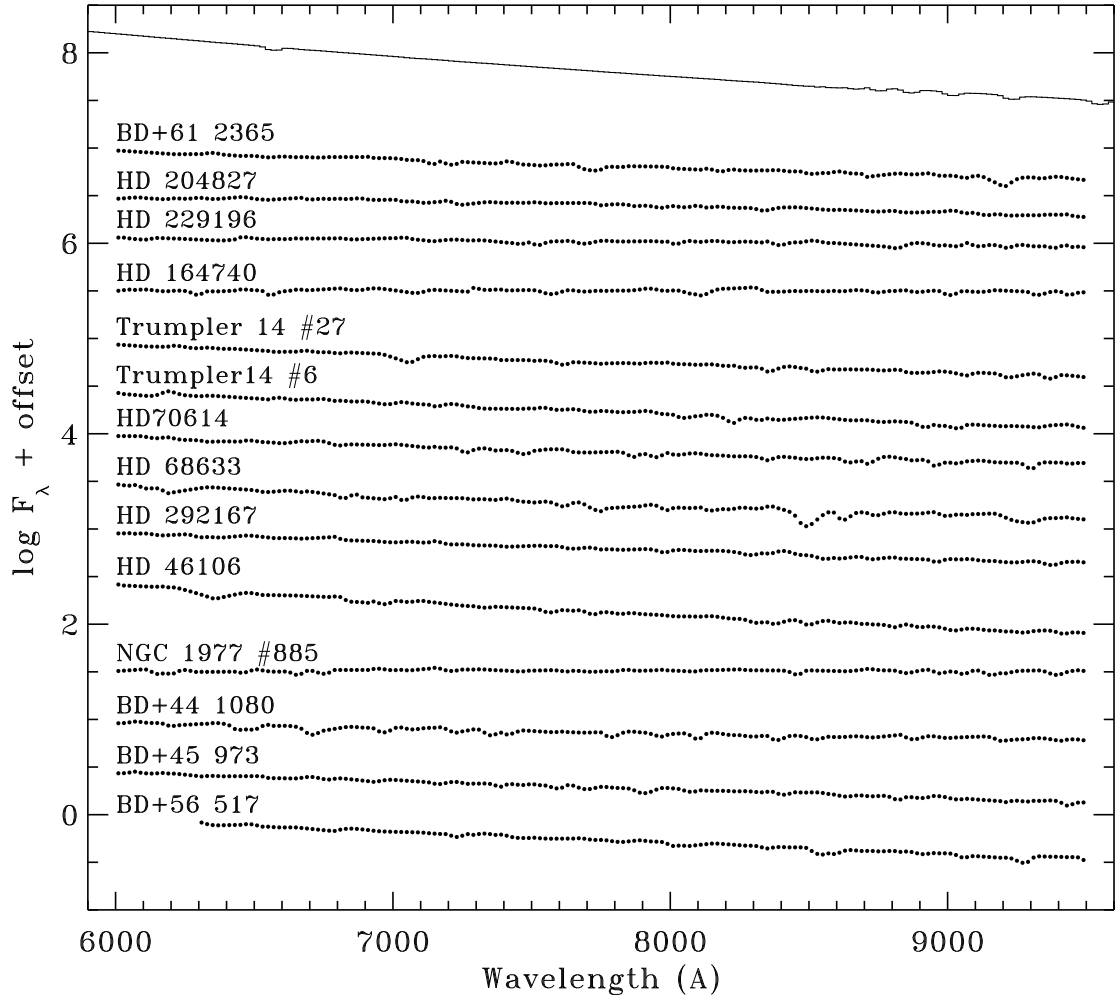}
\caption{Final processed ASC/HRC G800L spectra for the 14 program stars. A
Kurucz ATLAS9 model with \teff = 20000 K is shown near the top of the 
figure for comparison. Weak H$\alpha$ and H~I Paschen absorption lines may
be present in our spectra, although they would be close to the detection
limit. All the prominent ``bumps and wiggles'' seen in the ACS spectra are
instrumental in origin.
\label{fig_ACS}}
\end{figure}

The final set of 14 ACS/HRS/G800L spectra for our program stars are shown 
in Figure~\ref{fig_ACS}.  The spectra have been shifted vertically for 
clarity.  At the top of the figure, the SED of an ATLAS9 \citep{kurucz1991} 
model atmosphere with \teff\/ = 20000 K, \logg\/ = 4.0, and [m/H] = 0 is 
shown for comparison.  The model fluxes were computed in 20 \AA\/ bins in 
the spectral range shown and smoothed by 2-points to approximate the 
2-pixel resolution of the ACS data (i.e., $\sim$48 \AA).  While our program 
stars span a relatively wide range in \teff\/ ($\sim$15,000 -- 40,000 K), 
the model SED shown provides a reasonable depiction of the stellar features 
expected in our stars.  The H~I H$\alpha$ line at 6563 \AA\/ and the upper 
H~I Paschen lines near 9000 \AA\/ are the only spectral features of note 
for our targets and, at the resolution of the HRC/G800L, these features 
are very weak and generally undetectable.  An examination of the individual 
spectra show no signs of the H~I lines although numerous low amplitude
 ``bumps and 
wiggles'' are present.  These are instrumental features whose strength and 
location depend sensitively on the position of the stellar spectrum on the 
ACS detector.  In the discussions to come, these features are smoothed out 
or eliminated so as not to influence the analysis.

\subsection{Other Data\label{sec_OTHERDATA}}

As will be described below in \S~3, we will use the 
SED-fitting results from \citetalias{FMV}, and thus implicitly rely on the 
data used there.  These included low-resolution UV spectrophotometry from 
the \iue\/ satellite, obtained via the {Multimission Archive at STScI} 
(\mast); ground-based {\it UBV} photometry, obtained from the General 
Catalog of Photometric Data (GCPD) maintained at the University of Geneva 
(Mermilliod, Mermilliod, \& Hauck 1997)\footnotemark \footnotetext{The 
GCPD catalog was accessed at http://obswww.unige.ch/gcpd/gcpd.html.}; and 
{\it JHK} photometry from the {Two-Micron All Sky Survey} (\tmass), 
obtained from the \tmass\/ database at the NASA/IPAC Infrared Science 
Archive (\irsa)\footnotemark \footnotetext{The \tmass\/ data were accessed 
at http://irsa.ipac.caltech.edu/applications/Gator.}.  The only processing 
required for these data involved the \iue\/ spectrophotometry.  The \iue\/ 
data at \mast\/ were processed by the NEWSIPS software \citep{nichols1996}.  
As discussed in detail by \citet{massa2000}, these data contain significant 
thermal and temporal dependencies and suffer from an incorrect absolute 
calibration.  We corrected the data for their systematic errors and placed 
them onto the \hst/FOS flux scale of \citet{bohlin1996} using the 
correction algorithms described by \citet{massa2000}.  This step is 
essential for our program since our ``comparison stars'' used to derive 
extinction curves are stellar atmosphere models, so systematic errors in 
the absolute calibration would not cancel as they would for Pair Method of 
extinction determinations.

\subsection{A View of The Energy Distributions\label{sec_ALLTOGETHER}}

\begin{figure}[ht]
\figurenum{2}
\plotone{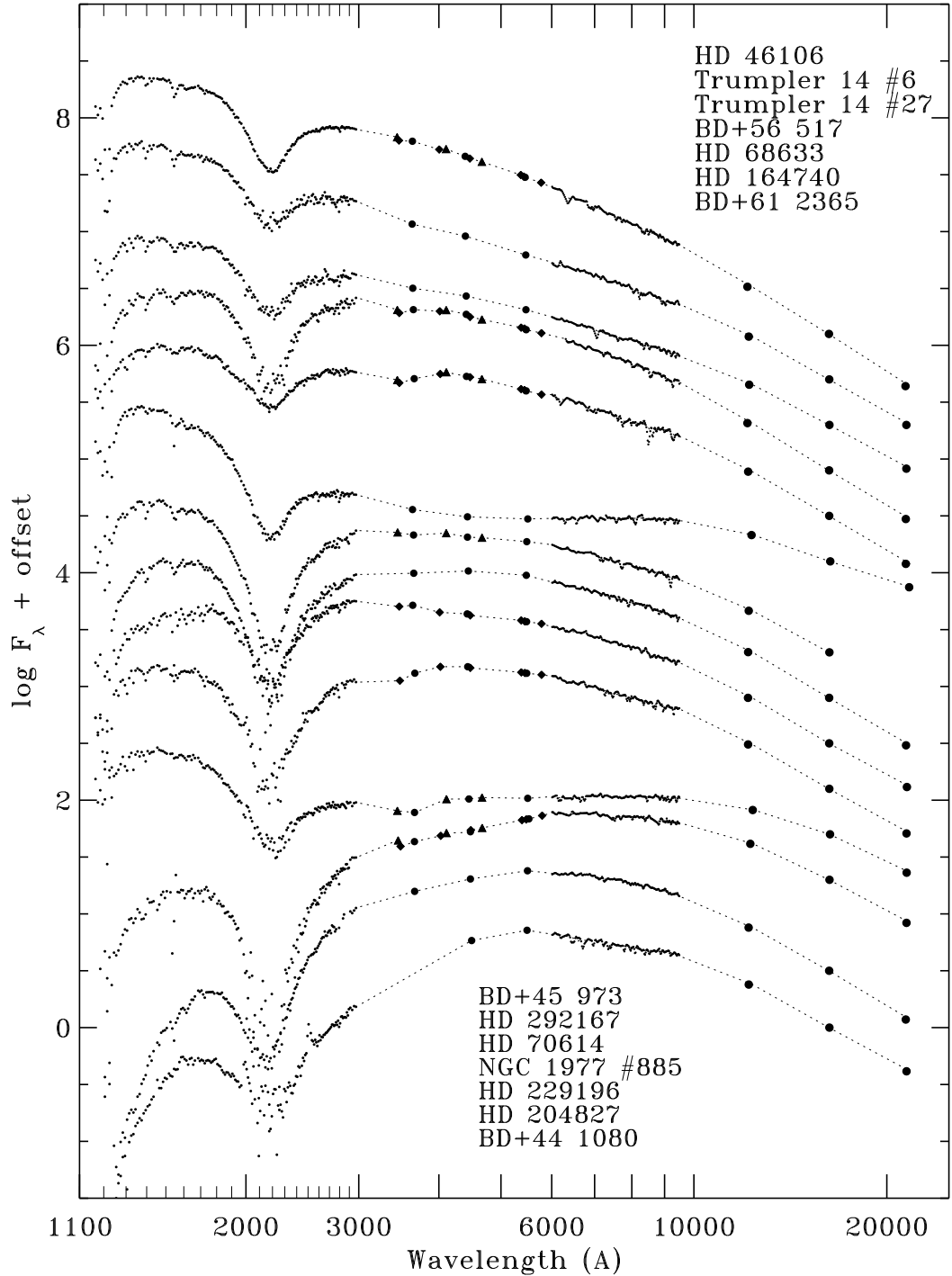}
\caption{Multiwavelength SEDs for the program stars. 
In the UV ($\lambda \leq 3000$ \AA), low-resolution \iue\/ spectrophotometry is 
shown as small circles. In the optical (3000 \AA\/ $< \lambda < 6000$ \AA), Johnson \ubv\/, Str\"{o}mgren 
{\it uvby}, and Geneva {\it UB$_1$B$_2$V$_1$G} photometry are indicated by the circles, triangles, and diamonds, respectively.  
In the NIR, the small circles are the new ACS observations and the 
large circles are the \tmass\/ \jhk\/ photometry.  The SEDs have been 
offset by arbitrary amounts for display purposes.  The dashed lines are not 
model fits to the SEDs, but were added to guide the eye and to help 
distinguish the datasets for each star.
\label{fig_SEDS}}
\end{figure}

To provide a perspective on the SEDs used to derive the extinction curves 
in our studies, Figure~\ref{fig_SEDS} shows all of the SED data for our 
14-star sample. The SEDs are arbitrarily shifted vertically for display.  
Small circles show the \iue\/ spectrophotometry in the range $\lambda \leq 
3000$ \AA, and the ACS/HRC/G800L data in the range $6000 \leq \lambda 
\leq 9500$ \AA. \tmass\/ \jhk\/ values are shown as large filled circles in 
the NIR region ($\lambda > 10000$ \AA). In the optical, \ubv\/ data are 
indicated as filled circles. To add to the coverage, we also show 
ground-based Str\"{o}mgren {\it uvby} (filled triangles) and Geneva {\it 
UB$_1$B$_2$V$_1$G} (filled diamonds) photometry from the GCPD catalog 
whenever they are available. For all the photometry, the conversion to 
absolute fluxes was performed using the calibrations of \citet{bstarsii}. 
As noted above, most of the structure seen in the ACS/HRC/G800L data is 
instrumental. The dashed curves in the figure are shown to guide the eye 
and to help isolate the individual stars. These lines are not model fits. 

\section{THE ANALYSIS\label{sec_ANALYSIS}}

We begin this section by creating the extinction curves that will be used 
in the analysis.  We then test whether the NIR extinction curves can 
be expressed as a simple power-law.  We demonstrate that there is no 
``universal'' exponent for the power law, and that the exponent which 
provides the best fit to a particular curve is sensitive to the wavelength 
range used to derive it.  We also show that the $R(V)$ determined from a 
power law fit to HD~164740 is inconsistent with previous estimates and with 
an independent estimate for $R(V)$\ based on its distance and absolute 
magnitude.  These facts destroy our confidence in schemes which attempt to 
derive $R(V)$\ from power law fits, and motivate us to introduce an 
alternative, analytic form for NIR extinction curves.  We show that 
this new form actually describes the NIR data as well as any power law, and 
that it does so over a much larger wavelength range.  Further, the new 
formulation produces an $R(V)$\ for HD~164740 which is consistent with the 
independent estimate.  It also gives $R(V)$\ values for the other program 
stars that agree with previous estimates.  

\subsection{THE NIR EXTINCTION CURVES\label{sec_CURVES}}

We produced normalized NIR extinction curves, $k(\lambda -V) \equiv 
E(\lambda -V)/E(B-V)$\ for our 14 sight lines by adopting the appropriate 
stellar atmosphere models derived in \citetalias{FMV} as the unreddened SEDs. 
In \citetalias{FMV} the observed SEDs of 328 reddened stars --- including 
those studied here --- were fit with a combination of stellar atmospheres 
(to estimate the unreddened SED) and a flexible analytical representation 
of interstellar extinction (to model $k(\lambda -V)$, utilizing a uniform 
set of \iue\/ UV spectrophotometry, \ubv\/ optical photometry, and \tmass\/ 
\jhk\/ NIR photometry.

For each program star here, we adopt the best-fit intrinsic SED from 
\citetalias{FMV}, which is defined by the values of \teff, \logg, [m/H], 
and $v_{turb}$ listed in their Table 3. The extinction in the \jhk\/ bands 
was computed using synthetic photometry of the stellar model fluxes as 
described in \citetalias{FMV} and references therein. To mitigate noise and 
more clearly view trends in the ASC data, we rebinned the ACS measurements 
and model SEDs into 500 \AA\ bins centered at 6250, 6750, 7250, 7750, 8250, 
8750, and 9250 \AA, thereby utilizing all of the ACS data within the 
6000--9500 \AA\/ range of our measurements.  These fluxes were then 
converted into 7 independent magnitudes.  To smooth instrumental features 
in the ACS data, we fitted the individual spectra with fourth-order 
polynomials, employing a sigma-clipping algorithm. The results were smooth 
representations of the ACS data, from which the synthetic magnitudes were 
computed.  We list the values of the normalized NIR extinction for the 
program stars in Table 2.  The labels ``{\it ACS62},'' 
``{\it ACS67},'' etc. refer to the seven synthetic filters described above.

The 1-$\sigma$ errors listed in the table incorporate both the uncertainty 
in the intrinsic SEDs and the uncertainties in the ACS, \jhk, and $V$-band 
measurements from which $k(\lambda-V)$\ is computed. For each star, we 
used the 100 Monte Carlo error simulations described in Paper V, which each 
yielded an independent estimate of the shape of the intrinsic SED.  For 
each of these SED's, we computed 20 NIR extinction curves, each using the 
observed photometry convolved with a random realization of the expected 
uncertainties in the measurements. Our noise model includes a 1-$\sigma$ 
error of $\pm$0.015 mag in $V$ and the standard errors provided with the 
\tmass\/ \jhk\/ photometry.  However, while calibrating synthetic 
photometry for use in SED modeling, we \citep{bstarsii} found that the 
\tmass\/ magnitudes -- notably $H$ -- had a larger scatter than could be
explained by the stated uncertainties.  We thus quadratically combine 
the standard \tmass\/ uncertainties with values of 0.007, 0.040, and 0.017
mag for $J$, $H$, and $K$, respectively, to arrive at the total uncertainty
for each magnitude.  Finally, we adopted the point-to-point photometric 
errors produced by the aXe processing software for the ACS/HRC/G800L 
spectra and a 1-$\sigma$ zero-point uncertainty of $\pm$2.5\% (see the 
Appendix). The final result was an ensemble of 2000 curves for each star, 
representing the likely range of uncertainties. The errors listed in 
Table 2 are simply the standard deviations of each point 
derived from the 2000-curve set.

The data from Table 2 are plotted in 
Figure~\ref{fig_CURVES}, with the key to the 
symbols given in the legend. Three curves are over plotted for most of the 
stars.  These are power-law model fits which will be discussed below.

\subsection{THE POWER-LAW MODEL\label{sec_MODEL}}

We begin by comparing our extinction curves to a basic power-law extinction 
model of the form
\begin{equation}
\label{eqnIR1}
A(\lambda) \; \propto \; \lambda^{-\beta} \;\; ,
\end{equation}
where $A(\lambda)$ is the total extinction at wavelength $\lambda$ and
$\beta$ is a wavelength-independent exponent. Converting this to the 
normalized form of our extinction curves yields
\begin{equation}
\label{eqnIR2}
k(\lambda -V) \equiv \frac{E(\lambda - V)}{E(B-V)} \; = \; k_{IR} 
\lambda^{-\beta} \; - \; R(V) \;\; ,
\end{equation}
where $k(\lambda-V)$\ is a function of the power $\beta$, a scale factor 
$k_{IR}$, and the well-known ratio of selective to total extinction $R(V)$ 
[$\equiv A(V)/E(B-V)$]. In the following discussion, we use non-linear 
$\chi^2$ minimization routines \citep{Markwardt2009}
to fit the 
observed extinction curves with equation~(\ref{eqnIR2}) for two different 
cases. In the first, 
$k_{IR}$ and $R(V)$ are taken as free parameters and a ``universal'' value
of $\beta = 1.84$ is assumed (the ``$\beta$-Fixed'' case). In the second, $k_{IR}$, $\beta$, and $R(V)$
are all allowed to vary (the ``$\beta$-Variable'' case).  Previous studies have shown that when extinction 
curves are modeled by equation~(\ref{eqnIR1}), the fits tend to diverge 
from observations at wavelengths shortward of the $I$ photometric band, 
i.e., at $\lambda \lesssim7500$ \AA\/ \citep[see, e.g., Figures 1, 2, and 
3 of][]{martin1990}.  Likewise, we found that, for most of our sight lines, 
the inclusion of ACS data shortward of 7500 \AA\/ greatly degraded the 
quality of the fits by increasing $\chi^2$.  Thus, we only use data with 
$\lambda > 7500$ \AA\ in our power law analyses.  These data are shown in 
Figure~\ref{fig_CURVES} as large filled symbols.

The best-fits to our data using equation~(\ref{eqnIR2}) are shown in 
Figure~\ref{fig_CURVES} as the dashed 
(``$\beta$--Fixed'') and solid (``$\beta$--Variable'') curves in each 
panel. The curves 
are extrapolated over the full wavelength extent of the panels, but were
determined using only the data indicated by the large filled symbols. In
most of the panels, it is very clear that the power-law functions fail
badly for $\lambda \lesssim 7500$ \AA\/ --- which led to the wavelength 
restriction described above.  The dotted curves shown in the Figures are 
$\beta$--Variable fits to the ACS data alone, utilizing all seven of the synthetic ACS magnitudes. These curves address the ability of a specific 
parameterization to provide a reasonable and consistent extrapolation to 
longer wavelengths. This important issue will be discussed further below.

Table 3 presents the numerical results of the fits
illustrated in Figure 3. The 1-$\sigma$ uncertainties listed in the table
are determined from the ensemble of 2000 Monte Carlo noise model
realizations computed for each sight line. For each parameter they represent
the dispersion of the results from power-law fits to the simulated curves
about the best-fit results. The final column of Table~3
gives the $F$-statistic, constructed from the $\chi^2$ values for the
$\beta$--Fixed and $\beta$--Variable cases. The $F$-statistic can be
used to determine whether the addition of a new fitting term (in this case
$\beta$) significantly improves the fit.  Large $F$ values, i.e., 
$F \gtrsim 10$, indicate the validity of the new parameter while 
progressively smaller $F$\ values indicate progressively less justification 
\citep[see, e.g.,][]{bevington1969}.  Although, to the eye, most of the 
fits in Figure~3 look reasonable, the $F$-statistic demonstrates that the 
fits for at least half of the sight lines are significantly improved by 
allowing a variable $\beta$, strongly justifying the inclusion of $\beta$ 
as a free parameter.  The largest $F$\ values tend to occur for the 
sight lines whose $R(V)$ values differ most from the Galactic mean of ~3.1, 
particularly those with the largest $R(V)$. The inability of the 
$\beta = 1.84$ model to adequately represent large $R(V)$\ sight lines is 
readily apparent in the last panel of Figure~\ref{fig_CURVES}.

An essentially identical method of assessing the statistical significance
of the derived $\beta$ values is simply to compare them with the
``universal'' value of 1.84. If the new values differ from 1.84 by
2-$\sigma$ or more --- which is the case for seven of our sight lines ---
then there is very strong statistical justification for rejecting 1.84 as
the appropriate value of $\beta$. A visual illustration of the significance
of the derived $\beta$'s is shown in the top panel of Figure
\ref{fig_BETA}, where we plot $\beta$ from the $\beta$--Variable case
against \ebv\/ for each sight line. The dotted line is $\beta = 1.84$. The 
sample mean and median values of $\beta$\ are 1.78 and 2.00, respectively.  
These are not far from the universal value, although the figure clearly 
illustrates significant departures exist. The figure also demonstrates that 
1.84 cannot be replaced by some better-determined universal value. 
Indeed, it challenges the notion of the existence of a universal IR 
extinction law.

A further challenge to the universal extinction model is shown in the lower
panel of Figure~\ref{fig_BETA}, where we plot $\beta$ against $R(V)$ for 
each sight line.  Note that these $R(V)$ values are determined by 
extrapolating the power-law models to $\lambda^{-1} = 0$ --- a practice 
which will be discussed further below. As might be anticipated from 
equation~(\ref{eqnIR2}), the uncertainties in $\beta$\ and $R(V)$ are 
correlated, and the directions and strengths of the correlations are 
indicated in the figure by the 1-$\sigma$ error bars. For comparison, we 
also show the values of $\beta$\ and $R(V)$ for the star $\rho$ Oph from 
\citet[filled triangle]{martin1990}. The consistency of the power-law 
exponent for this dense cloud sight line with the results for diffuse cloud 
sight lines has been a significant motivation for the perception of a 
universal $\beta \simeq 1.8$\ law \citep[e.g., Fig.~3.7 in][]{whittet2003}. 
Although the $\rho$ Oph measurements were not made in the same manner as 
ours, they utilize data over the same wavelength range ($I$ through 
$K$-band) and should be compatible. The data in Figure \ref{fig_BETA} 
reveal a striking systematic trend --- $\beta$ is anti-correlated with 
$R(V)$ in the sense that large-$R(V)$ sight lines have smaller than average 
values of $\beta$. As $R(V)$ is generally considered a coarse indicator of 
grain size, i.e., populations of larger-than-average grains yield 
larger-than-average $R(V)$ \citep[e.g., see][]{mathis1981}, our results indicate 
that the shape of IR extinction curves parameterized by $\beta$\ are
environmentally-dependent.  As shown in Paper V, there is considerable
intrinsic scatter in correlations between $R(V)$ and other properties of
interstellar extinction curves, and often the correlations are only
apparent when the most extreme $R(V)$\ values are considered. Our results
suggest that previous analyses have not had access to a wide-enough range
in $R(V)$ to reveal the environmental dependence of $\beta$.

Since a major motivation for using an analytic form for extinction is to 
extrapolate the curve to infinite wavelength and estimate $R(V)$, it is 
troubling that the $R(V)$s derived from the power law curves depend 
strongly on the adopted exponent.  Figure 3 and Table 3 show 
that the $\beta$--Fixed and $\beta$--Variable approaches give similar 
results when $R(V)$\ is near the Galactic mean of $\sim$3.1, but yield 
increasingly divergent results at both larger and smaller $R(V)$.  This is 
best illustrated by the HD~164740 sight line.  This is the most extreme 
sight line in our sample in the sense that it has the smallest $\beta$\ and 
largest $R(V)$.  Furthermore, there is no indication that it is 
pathological, since Figure \ref{fig_BETA} shows that it is consistent with 
an extrapolation of the trends exhibited by less extreme sight lines. The 
problem with HD~164740 is that, even though the $\beta$--Variable fit 
provides a fine representation for the $I$\ through $K$-band extinction 
(see Figure~\ref{fig_CURVES}), the extrapolation of its power-law fit to 
longer wavelengths is inconsistent with the physical expectation that 
$\frac{d k(\lambda -V)}{d \lambda} \rightarrow 0$ \ as $\lambda \rightarrow 
\infty$, (which demands $\beta \ge 1$).  
 
An additional, but related, problem is that the $R(V)$\ determined from the 
$\beta$--Variable fit for HD~164740 is demonstrably incorrect.  HD~164740 
is a member of the cluster NGC~6530, whose distance, $d$, has been 
determined independently of HD~164740.  Therefore, given its absolute 
magnitude, $M_V$, it is possible to derive an independent estimate for 
$R(V)$\ by rearranging the formula for the distance modulus.  Specifically, 
\begin{equation}
\label{eqnDM2}
R(V) = \frac{V - M_V - 5 \log(d/10 pc)}{E(B-V)} \;\; ,
\end{equation}
The data needed to compute $R(V)$ for HD~164740 are $d = 1330$~pc 
(Paper~V), $M_V = -4.8$ \citep{Walborn1973}, $V$ and $E(B-V)$ (see 
Table~1).  We consider two cases which span the range of 
expected absolute magnitudes: a single star or a binary with identical 
components.  Using these data for HD~164740, we find $5.2 \leq R(V) \leq 
6.1$.  These values are significantly smaller than the $R(V) = 7.0$\ 
inferred from extrapolation of the $\beta$--Variable power-law fit.  One 
can appeal to uncertainties in the distance and absolute magnitude of 
HD~164740, but these are insufficient to explain such a large discrepancy.  
In support of this conclusion, \citet{hecht1982} find $R(V) = 5.6$ based 
on a comparison of the brightness and color of HD~164740 with those of the 
nearby -- and presumably equally distant --- O-type star HD~164794 (9~Sgr). 
Note that other estimates of $R(V)$ for HD~164740 have been made by 
extrapolating an assumed form for NIR extinction by, for example, 
\citet{CCM} [$R(V) = 5.30$] and more recently by us in Paper~V [$R(V) = 
5.21$]. However, the assumed forms, which are (or resemble) power-laws 
with large exponents ($\beta = $1.6--1.84) are clearly inconsistent with 
the observed shape of the NIR extinction curve of HD~164740.  Thus, 
these results cannot be used to prove an error in the $\beta$--Variable 
value.

The HD~164740 discussion shows that, even though a power-law can fit an 
NIR extinction curve quite well over a specific wavelength interval, 
extrapolation of such fits to longer wavelengths can be unreliable. This 
was anticipated by \citet{martin1990} who noted that it would be 
``remarkable'' for a single power-law to be appropriate over an extended 
wavelength range.  Nevertheless, this caveat is explicitly ignored when a 
single-exponent power-law is used to estimate $R(V)$.  To emphasize this 
point, consider the dotted curves in Figure~3.  These are power-law 
fits (eq.~[\ref{eqnIR2}]) to all of the ACS data, i.e., covering the full 6000 -- 9500 \AA\/ wavelength range.  While they certainly provide an excellent analytical representation of $k(\lambda -V)$ at these wavelengths ($\chi^2 < 0.2$ and, in 12-of-14 cases, $\chi^2 < 0.1$), nevertheless it is quite clear that extrapolations of these fits to longer wavelengths are grossly inconsistent with the \jhk\/ data and, consequently, the intercepts of the extrapolated 
curves are totally unrelated to $R(V)$.  In addition, 
the $\beta$s are always smaller than those inferred from the $I$ through 
$K$-band fits in Table 3 and are nearly always less than the 
limiting value of one.  These results suggest a systematic effect in which the exponent of a best-fit power-law systematically decreases as shorter wavelengths are considered and the implied $R(V)$ increases.

At this point, several straightforward conclusions can be drawn. 
\begin{enumerate}
\item All of our NIR extinction curves can indeed be well-represented by a 
simple power-law formulation in the $I$-band through $K$ band spectral 
region. 

\item However, the power-law exponent, $\beta$, varies significantly from 
sight line-to-sight line, and is incompatible with the thesis of a 
universal form to NIR extinction.  

\item The value of the exponent, $\beta$, is anti-correlated with $R(V)$, 
suggesting that the detailed shape of NIR extinction curves is likely a 
function of the line of sight grain size distribution.  

\item The parameters of power law fits can depend quite strongly on the 
wavelength interval used to determine them, with the value of $\beta$\ 
tending to decrease as shorter wavelengths are included.  This result 
implies that extrapolations of power law fits to longer wavelengths can be 
an unreliable means for estimating $R(V)$.  

\end{enumerate}

\subsection{AN ALTERNATIVE MODEL\label{sec_PEI}}
Given the problems encountered with the simple power law representation 
for NIR extinction, we decided to search for a different functional form, 
whose parameters can be estimated from accessible data and can be 
extrapolated to long wavelengths to obtain consistent, meaningful estimates 
of $R(V)$.  While many such forms are possible, the one introduced by 
\citet{Pei1992} in a study of Milky Way and Magellanic Cloud extinction 
produced remarkably good results\footnote{We thank Karl Gordon for bringing 
this form to our attention. It was used in a study of Galactic extinction curves by \citet{clayton2003b} and a more general discussion of its utility will be given in \citet{Gordon2009}}.  We use a generalized version of 
the function, with more degrees of freedom.  It has the form 
\begin{equation}
\label{eqnPEI1}
k(\lambda - V) \; = \; k_{IR} 
\frac{1}{1+(\lambda/\lambda_0)^{\alpha}} \; - \; R(V) \;\; ,
\end{equation}
where $\alpha$, $k_{IR}$, $\lambda_0$ and $R(V)$ are potentially free 
parameters. For $\lambda \gg \lambda_0$, this function reduces to a power 
law with exponent $-\alpha$.  For $\lambda \ll \lambda_0$, it flattens to a 
constant whose value is $k_{IR} - R(V)$.  Between these extremes, it 
resembles power law whose exponent increases with wavelength --- which is 
exactly the behavior suggested by our analysis in the previous section.

While the introduction of four free parameters may seem a steep price to 
pay for abandoning the simpler power-law formulation, we quickly discovered 
that two strong observational constraints can be placed on the parameters.  The 
first is that $\lambda_0$\ can be replaced by constant for all of our sight 
lines, which span a wide range of extinction properties, without 
significant loss of accuracy.  The second is that the scale factor, 
$k_{IR}$, can be expressed as a linear function of $R(V)$, again without 
loss of accuracy.  These constraints result in a simplified version of 
equation~(\ref{eqnPEI1}) given by 
\begin{equation}
\label{eqnPEI2}
k(\lambda - V) \; = \; [0.349 + 2.087R(V)] 
\frac{1}{1+(\lambda/0.507)^{\alpha}} \; - \; R(V) \;\; ,
\end{equation}
where $\lambda$ is in $\mu$m.  Equation~(\ref{eqnPEI2}) contains only two 
free parameters, $\alpha$ and $R(V)$; the same number as the $\beta$--Fixed 
power law model, and one less than the $\beta$--Variable model.  

Equation~(\ref{eqnPEI2}) proved so successful in reproducing the shapes of 
the NIR extinction curves that we applied it to the \jhk\/ data and the 
entire wavelength range of the ACS data.  The results are given in 
Table~4 and Figures~5 and \ref{fig_BETA_PEI}. In Figure~5, the 
solid portion of the curves indicate the wavelength interval over which the 
fit was performed, and the dotted portions show extrapolations of the fit 
to longer and shorter wavelengths.  Note that, although the $V$-band point 
was not included in the fits, extrapolations of them pass quite close to it.  
In fact, the $V$-band could have been included in the fits without much 
loss of overall accuracy.  Table 4 gives the $\alpha$\ and 
$R(V)$ that result from the fits and their uncertainties.  These are 
plotted in Figure~\ref{fig_BETA_PEI}. The correlated uncertainties were 
computed as described in \S~\ref{sec_MODEL}.  The range in $\alpha$\ and 
the relationship between $\alpha$\ and $R(V)$ mimic those of the 
$\beta$--Variable fits discussed in \S~\ref{sec_MODEL}.

That equation~(\ref{eqnPEI2}) provides a much better fit to the observed 
NIR extinction curves can be seen by comparing the mean $\chi^2$ values 
listed in Tables~3 and 4 for the different 
models.  The mean $\chi^2$ values for the $\beta$--Fixed, and 
equation~(\ref{eqnPEI2}) are 0.78 and 0.24, respectively.  Thus, even 
though the fits using equation~(\ref{eqnPEI2}) include more points, their 
mean $\chi^2$ is much better than the 2-parameter $\beta$--Fixed power 
law model.  Furthermore, unlike the $\beta$--Fixed model, 
equation~(\ref{eqnPEI2}) provides a good fit for {\em all} of the curves.  

Notice that although the mean $\chi^2$\ for the 3-parameter 
$\beta$--Variable from Table~3 is slightly better than the 
equation~(\ref{eqnPEI2}) value (0.17 compared to 0.24), a fairer comparison 
between the two is given in Table~4.  The last column of the 
table lists the $\chi^2$\ values that result from fitting $\beta$--Variable 
models to the same data as the equation~(\ref{eqnPEI2}) fits.  In this case, 
we see that the mean $\chi^2$ for the $\beta$--Variable fits is 0.44, 
considerably larger than the 0.24 that results from the 
equation~(\ref{eqnPEI2}) fits.

An additional advantage to the formulation in Equation~(\ref{eqnPEI2}) is its robustness to changes in the wavelength coverage of the data.  For instance, if we fit only the ACS data, spanning the wavelength range 6000 -- 9500 \AA, then the resultant values of $R(V)$ differ on average by only +0.09 from the results in Table 4, with a mean scatter of $\pm$0.27 and no systematic trends.  Likewise the implied $\alpha$'s differ on average from those in Table 4 by $-0.05$, with a scatter of $\pm$0.20.  Thus, in striking contrast to the $\beta$-Variable case, a reliable and stable estimate of IR extinction can be derived, even in the absence of longer wavelength \jhk\/ data. 	

It is important to determine whether the values of $R(V)$\ determined by 
equation~(\ref{eqnPEI2}) agree with previous estimates.  To begin, we note 
that the $R(V)$ value determined for HD~164740 is 6.00, within the expected 
range described in \S~\ref{sec_MODEL}.   We can also make a broader 
comparison by comparing the values in Table~4 to a commonly used 
means to estimate $R(V)$, which is based on optical and JHK photometry.  
This approach estimates $R(V)$ from the relation $R(V) = -1.1 \frac{E(K-V)}{E(B-V)}$, 
which is based on van de Hulst's theoretical extinction curve No. 15 
\citep[e.g.,][]{Johnson1968}.  While, in general, our results show that two 
parameters are needed to fully define an NIR extinction curve, this does 
not exclude the possibility that relationships may exist between specific 
properties of the curves, at least to a high degree of approximation.  
Figure~\ref{fig_KLAMK} is a plot of $\frac{E(K-V)}{E(B-V)}$ (Table~2) 
versus $R(V)$ from Table~4.  The two quantities are clearly 
strongly correlated.  The van de Hulst--based formula (dashed line) 
follows the general trend for low and moderate values of $R(V)$, but 
systematically underestimates it for large-$R(V)$ sight lines.  
Nevertheless, the overall agreement is quite good, with an RMS deviation of 
0.26 in $R(V)$, implying that the $R(V)$ values determined by 
equation~(\ref{eqnPEI2}) are in general agreement with previous estimates.  

In fact, if we assume that the $R(V)$ derived from equation~(\ref{eqnPEI2}) 
are correct, we can derive a more accurate means for determining $R(V)$\ 
from the broad band photometry.  The solid line in Figure~\ref{fig_KLAMK} is 
the unweighted linear fit, $R(V) = -1.36 \frac{E(K-V)}{E(B-V)} - 0.79$, which has an RMS 
deviation for $R(V)$ of 0.13.  Note that both lines yield similar results 
for $R(V) \lesssim 3.5$.

The major conclusions from this section are:
\begin{enumerate}
\item Equation~(\ref{eqnPEI1}) mimics a power law whose exponent increases 
with wavelength, which is exactly the behavior our analysis in 
\S~\ref{sec_MODEL} implied was needed.  

\item Initial fits to the data demonstrated that the number of free 
parameters needed to fit NIR curves can be reduced to 2 (see, 
eq.~[\ref{eqnPEI2}]).

\item Equation~(\ref{eqnPEI2}) provides good fits to all of the extinction 
curves, and even performs better than the $\beta$--Variable model, which 
has one more free parameter.  

\item Equation~(\ref{eqnPEI2}) produces $R(V)$\ values which are consistent 
with independent estimates, and are in accord with commonly used methods 
for estimating $R(V)$.  

\item Equation~(\ref{eqnPEI2}) gives $R(V)$\ values that do not depend 
strongly or systematically on the wavelength interval used to derive them.  

\item For our data, the relation  $R(V) = -1.36 \frac{E(K-V)}{E(B-V)} - 0.79$ reproduces 
$R(V)$ to within $\pm 0.12$ over a large range in $R(V)$. 

\end{enumerate}

\section{DISCUSSION\label{sec_DISCUSS}}

In general, there are two motivations for determining an analytical model 
for the shapes of interstellar extinction curves.  The first is that it 
allows a consistent interpolation or extrapolation of extinction properties 
into spectral regions where measurements are lacking.  This is used to 
define $R(V)$ and, hence, $A(V)$.  The second is that the functional 
dependence of extinction over some particular spectral region may shed 
light on the physical processes controlling the extinction, e.g., the 
composition and size distribution of the grains responsible for the extinction.  Further, the 
analytic representation may yield specific parameters whose range of 
variability could be related to how the grain populations respond to 
different physical environments along the line of sight.  The first of 
these goals was addressed in the previous section.  Here, we focus on the 
second.  

Our results suggest that at least two parameters are needed to characterize 
the NIR extinction, $R(V)$\ and $\alpha$.  It seems reasonable to suggest 
that $R(V)$ provides a measure of the mean grain size along the line of 
sight, and that $\alpha$\ is probably related to the grain 
size distribution.  This is because, in the small particle limit (where the 
ratio of grain-size to wavelength is small -- as in the IR), Mie scattering 
theory implies that both absorption and scattering are expected to approach 
power law forms with an exponent of 4 for scattering and $\sim$1 for 
absorption \citep[see Chapter 7 of][]{Spitzer1978}. Power-law exponents 
determined in the NIR generally lie between these two extremes. Given the 
sensitivity of extinction to grain properties, however, it would be 
surprising for extinction to adhere strictly to a single-exponent power law 
form over an extended wavelength range \citep{martin1990}.

In fact, these arguments make the suggestion of a universal power-law
exponent for the IR troubling.  It is clear from the study of other aspects of 
extinction (e.g., the shape of UV extinction curves) that the properties of 
grain populations differ significantly from place to place in interstellar 
space. As \citet{martin1990} note, since the absorption and scattering 
properties of grains are generally strongly dependent on the grain 
composition and size, the suggestion of universality requires highly 
constrained sight line-to-sight line modifications in grain properties 
which must occur in such a way as to yield, for example, a range in $R(V)$ 
but a single IR power-law exponent.  Our results remove this troubling 
constraint.  We find that the exponent is spatially variable and 
likely does reflect underlying differences in the grain populations. 

To verify that the observational parameters obtained from 
equation~(\ref{eqnPEI2}) are consistent with grain models and to determine 
whether they can be used to constrain the grain parameters, we compare our 
results with the theoretical calculations given by \citet[][hereafter
MW81]{mathis1981}.  Although more recent and sophisticated models exist, 
the \citetalias{mathis1981} models are simple, accessible, and adequate to
demonstrate the points we wish to make. In their Table~1, 
\citetalias{mathis1981} present the extinction optical depths (per
$10^{22}$ H atoms) for separate populations of silicate and graphite
grains, each with a power-law size distribution \citep[following the work
of][hereafter MRN]{MRN} and six different combinations of the upper and
lower grain-size cutoffs, $a_+$ and $a_-$, respectively. We produced model
extinction curves by summing the graphite and silicate extinction for all
possible combinations of the six size distributions (of which there are 36)
and three different assumed values for the dust-phase C/Si ratio. As a
baseline, we adopt a ratio (C/Si)$_{dust} = 6.5$, which is consistent with
the mean for the variety of dust grain models considered by
\citet{zubko2004} and also with mean dust-phase abundances based on solar
composition \citep[see, e.g., Table 2.2 of][]{whittet2003}. We used this
value for one set of 36 model optical depth curves and values a factor-of-2
lower (3.25) and higher (13) to produce two additional sets. After
constructing this ensemble of 108 curves, we converted them from optical
depth to absolute extinction and then normalized them by $E(B-V)$,
producing curves of $A(\lambda)/E(B-V)$. The justification for producing
this multitude of curves is simple: we have no {\it a priori}\ reason to 
believe that the
grain-size distributions must be identical for the graphite and silicate
grain populations, as the destruction and, certainly, the growth mechanisms
are unlikely to be identical. Furthermore, we do not know the exact
abundances of C and Si in the ISM, and suspect that the dust-phase C/Si ratio varies from
sight line--to--sight line. Thus our model curves represent a wide range of
conditions that may occur in interstellar space.

We seek to compare the shapes of model IR curves with specific $R(V)$\ 
values to the observed parameters determined from equation~(\ref{eqnPEI2}). 
Since the \citetalias{mathis1981} curves are presented at only at a small 
number of wavelength points, including optical/IR points at $B$, $V$, 1.11 
\invmic\/ and 0.29 \invmic\/, we take $R(V)$ as the value of 
$A(\lambda)/E(B-V)$ at the $V$ point and characterize the shape of the IR 
with a power law fitted through the points at 1.11 \invmic\/ ($\lambda$ = 
0.9 \mic) and 0.29 \invmic\/ ($\lambda$ = 3.4 \mic), yielding an exponent 
we refer to as ``$\alpha_{MW81}$.'' This provides a measure of the general 
curvature of the extinction between 0 and 1.11 \invmic\/ (i.e., $\lambda > 
9000$\AA). A fair comparison with the predictions of 
equation~(\ref{eqnPEI2}) requires that we compute a similar value of 
$\alpha_{MW81}$ for each of the curves in Figure~5, using 
equation~(\ref{eqnPEI2}) at the same wavelengths.

The comparison between $\alpha_{MW81}$ and $R(V)$ for the models and the
observations is shown in Figure~\ref{fig_BETA_PEI_2}. Filled circles show 
the observational results.  The distribution of points is similar to that 
in Fig.~\ref{fig_BETA_PEI}, except that the values of $\alpha_{MW81}$ 
are $\sim 0.15$--0.20 smaller than $\alpha$.  Measurement errors were 
determined in the same way as described above.  The theoretical 
$\alpha_{MW81}$ and $R(V)$ values derived from the 108 model curves are 
depicted as the x's.  The symbol sizes correspond to the (C/Si) ratio in 
the dust, with (C/Si)$_{dust} = 3.25$, 6.5 and 13 represented by small, 
medium, and large crosses, respectively.  Two interesting points are 
evident in the Figure.  First, the main distribution of model points at 
large $R(V)$ aligns nicely with the ``observed'' values of $R(V)$ and 
$\alpha_{MW81}$.  This indicates that the functional form proposed in 
equation~(\ref{eqnPEI2}) is compatible with theoretical calculations.  
Second, there is a group of model points that do not overlap the domain 
covered by the observations (the models which have small $\alpha_{MW81}$ 
{\em and} small $R(V)$ values).  This may indicate that there are specific 
size distribution and abundance combinations which are not favored by 
nature.  In particular, most of these points arise from the combination 
of a silicate grain population with a small upper size cutoff ($a_+ = 
0.25$) and a graphite grain population with a large upper size cutoff 
($a_+ = 0.40$ or 0.50).  Note that, because \citetalias{mathis1981} were 
explicitly interested in studying the behavior of curves with large 
$R(V)$ values, they considered grain size distributions with larger mean 
sizes than the \citetalias{MRN} distribution, which provides a reasonable 
fit for the Galactic mean curve with $R(V) = 3.1$.  As a result, there 
are no models with large $\beta$\ and small $R(V)$ to compare to the 
observations.  Additional calculations and more sophisticated grain
models, such as those discussed by \citet{zubko2004}, will be needed to
address this.  However, it is the large $R(V)$ sight lines where our
results differ most strongly from the paradigm of a universal power law 
and, in these cases, the comparison in Figure \ref{fig_BETA_PEI_2} is 
quite satisfactory.

From the preceding discussion, we conclude that the two parameters needed 
to fit the observations using equation~(\ref{eqnPEI2}) are probably the 
minimum required.  This means that the hope of finding a one parameter 
representation, such as a universal power law with an adjustable $R(V)$\ 
or a model of the sort described by \citet{CCM}, where the curve shape 
can also have an $R(V)$\ dependence, are probably unjustified.  We were 
also able to demonstrate how examining the observed parameters can 
provide useful constraints on the physical parameters of dust models.  

\section{SUMMARY AND FINAL COMMENTS\label{sec_SUMMARY}}

Collecting the results from the previous sections, we have drawn the 
following conclusions from our analysis of NIR interstellar extinction 
along 14 sight lines:

\begin{enumerate}
\item Our NIR extinction curves can be well-represented by a power-law 
formulation in the $I$-band through $K$ band spectral region.  However, 
the exponent of the power-law varies significantly from sight 
line-to-sight line, which is incompatible with the notion of a universal 
form to NIR extinction.  

\item The value of exponent in the power law, $\beta$, is anti-correlated 
with $R(V)$, showing that the shape of NIR extinction curves are likely 
functions of the line of sight grain size distributions.  

\item The parameters derived from a power law fit can depend quite strongly 
on the wavelength interval used to determine them, with the value of 
$\beta$\ inferred from the fit increasing as longer wavelengths are 
included.  This result demonstrates that extrapolations of power law fits 
to longer wavelengths are an unreliable means for estimating $R(V)$.  

\item Because of the biases in the power law, we have adopted a different 
form for the NIR extinction, equation~(\ref{eqnPEI1}), which has the 
following properties:

\begin{itemize}
  \item Equation~(\ref{eqnPEI1}) mimics a power law whose exponent increases 
  with wavelength -- a property demanded by the power law analysis.  

  \item The 4 free parameters in equation~(\ref{eqnPEI1}) can be reduced to 2 
  (see, eq.~[\ref{eqnPEI2}]), without affecting the quality of the fits to 
  the NIR extinction curves.

  \item For $R(V) \sim 3$, the curve produced by equation~(\ref{eqnPEI2}) is
  very similar to the commonly used $\beta = 1.84$\ power law.

  \item Equation~(\ref{eqnPEI2}), with its 2 free parameters, provides better 
  fits to all of the extinction curves than the power law model which allows 
  the exponent to be a free parameter -- giving it 3 free parameters.  

  \item Equation~(\ref{eqnPEI2}) produces $R(V)$\ values which are consistent 
  with independent estimates, and are in accord with commonly used methods 
  for estimating $R(V)$.  

  \item Unlike the power law models, equation~(\ref{eqnPEI2}) gives $R(V)$\ 
  values that do not depend strongly or systematically on the wavelength 
  interval used to derive them.  
\end{itemize}

\item For our data, the simple relation $R(V) = -1.36 \frac{E(K-V)}{E(B-V)} - 0.79$ 
reproduces the $R(V)$ determined from equation~(\ref{eqnPEI2}) to within 
$\pm 0.12$, over a large range in $R(V)$.  This is a useful relation for estimating $R(V)$ when only limited broad band photometry is available.  

\item Using theoretical grain models, we showed that the two parameters 
needed to fit the observed curves using equation~(\ref{eqnPEI2}) are 
probably the minimum required.  As a consequence, the notion of finding a 
one parameter representation, such as a universal power law with an 
adjustable $R(V)$\ or a \citet{CCM}-like model where the curve shape can 
also have an $R(V)$\ dependence, are probably unjustified.  

\item We demonstrated how examining the observed parameters can provide 
useful constraints on the physical parameters of dust models.  

\end{enumerate}

Our results bear heavily on the recent observations of ``non-universal''$\beta$ values noted in \S~1 \citep[i.e.,][]{larson2005, nishiyama2006, froebrich2007, gosling2009} and indicate that they should be viewed with some caution.  On the one hand, these may reflect the intrinsic variability expected from an interstellar medium with a wide range in physical properties and grain processing histories for the specific conditions along the respective sight lines.  However, care must be taken in comparing $\beta$ values from different datasets. Our results show that power-law models for a given sight line will yield different exponents, depending on the wavelength range used to derive them.  For example, a $\beta$-Variable power-law fit to equation~\ref{eqnPEI2} in the \jhk\/ region gives exponents that are $\sim$0.3 larger than those determined by fits which also include the $I$-band. This effect is likely present to some degree in the above studies, which are based solely on \jhk \/ data, and should be taken into account before we can obtain an accurate picture of the range 
in IR extinction properties.

Finally, we point out that while our results are interesting, they are 
based on only 14 sight lines.  Clearly, more NIR data are needed to verify 
the generality of the constraints placed on equation~(\ref{eqnPEI1}) to 
obtain equation~(\ref{eqnPEI2}), and to verify the relationship 
$R(V) = -1.36 \frac{E(K-V)}{E(B-V)} - 0.79$.

\begin{acknowledgments}
We thank the referee, G. Clayton, for raising issues that prompted us to 
reconsider certain aspects of our initial draft and, thereby, improve and 
extend our results.  E.F. acknowledges support from NASA grant 
HST-GO-10547.01-A.  D.M. acknowledges support from NASA grant 
HST-GO-10547.02-A. Some of the data presented in this paper were obtained 
from the Multimission Archive at the Space Telescope Science Institute 
(MAST). STScI is operated by the Association of Universities for Research 
in Astronomy, Inc., under NASA contract NAS5-26555. Support for MAST for 
non-HST data is provided by the NASA Office of Space Science via grant 
NAG5-7584 and by other grants and contracts. This publication also makes 
use of data products from the Two Micron All Sky Survey, which is a joint 
project of the University of Massachusetts and the Infrared Processing and 
Analysis Center/California Institute of Technology, funded by the National 
Aeronautics and Space Administration and the National Science Foundation.
\end{acknowledgments}

\appendix
\section{Absolute Calibration of the ACS/HRC/G800L Data}

Because the absolute calibration is critical to our program, we took the 
additional step to verify the HRC/G800L calibration by processing G800L 
spectra for three spectrophotometric standard DA white dwarfs (G191B2B, 
GD~71, and GD~153) used to help establish the fundamental {\it HST}\ flux 
calibration \citep[see, e.g.,][]{bohlin1996}. Our procedure was simple. We 
downloaded ACS/HRC/G800L spectra for these stars from MAST, processed them 
in exactly the same manner as our target stars, and then compared the 
resultant spectra with the model fluxes used to establish the absolute 
calibration. The models were obtained from the CALSPEC Calibration Database 
at the Space Telescope Science Institute\footnotemark 
\footnotetext{http://www.stsci.edu/hst/observatory/cdbs/calspec.html} 
(the files g191b2b\_mod\_004.fits, gd71\_mod\_005.fits, and 
gd153\_mod\_004.fits).

Figure~\ref{fig_CALSPEC} shows the results of this comparison. The 
different symbols show the ratio of the model fluxes to the ACS fluxes as a
function of wavelength for each of the three stars. Clearly, the processed 
HRC/G800L fluxes are increasingly underestimated at wavelengths longward of 
$\sim$6000 \AA, with the effect increasing dramatically below $\sim$9500 
\AA. Fortunately, the flux discrepancy is systematic and consistent among 
the three standards and, thus, can be corrected in a simple way. We derived 
a correction curve for our processed spectra by fitting a polynomial to the
data in Figure~\ref{fig_CALSPEC} and the result is shown as the smooth
solid curve in the figure. The systematic scatter of the three stars about
the mean correction suggests a zero point stability in the fluxes of
$\sim$2.5\% at wavelengths shortward of 9500 \AA. The variability increases
significantly at longer wavelengths. The mean correction in 
Figure~\ref{fig_CALSPEC} is applied as a multiplicative factor to the
calibrated results of the aXe processing.

To perform an independent test of our calibration correction, we examined
its effect on the HRC/G800L spectrum of the Galactic O7 star HD~47839
(15~Mon). ACS spectra for HD~47839 were obtained from MAST, and processed
in the same way as our program and spectrophotometric calibration stars.
Figure~\ref{fig_15MON} shows the multi-wavelength SED of HD~47839 spanning
the range $\sim$1150 \AA\/ to $\sim$2.2 $\mu$m. In the main figure are
\iue\/ low-resolution UV spectrophotometry ($\lambda \leq 3000$ \AA),
Johnson, Str\"{o}mgren, and Geneva optical photometry (3300 \/$< \lambda <$
6500 \AA), and Johnson {\it RI} and 2MASS {\it JHK} NIR photometry
($\lambda >$ 6500 \AA). The solid curve is a 37,000 K TLUSTY stellar
atmosphere model \citep{lanz2003} fitted to the data in the manner
described in our previous papers \citep[e.g.,][]{FMIV, bstarsii}. The inset
to the figure illustrates the effect of our HRC/G800L correction curve. The
smooth curve shows the best fit model while the corrected and uncorrected
G800L data are over plotted. The HD~47839 data clearly confirm the results
from the analysis of the spectrophotometric standards and the effectiveness
of the calibration correction in Figure~\ref{fig_CALSPEC}. At wavelengths
shortward of 9500 \AA\/ the wavelength-dependent signature of the
calibration deficiency has been removed and the mean flux level lies about
1\% above that predicted by the best-fit SED model, consistent with the
scatter among the spectrophotometric standards. At wavelengths longward of
9500 \AA, the fluxes are clearly much less reliable.

The calibration correction curve derived in Figure~\ref{fig_CALSPEC} is
available from the authors, although it should be noted that is applicable
only for first-order HRC/G800L data which are processed using the same aXe
parameters as we adopted for our processing, particularly, an extraction
window height of $\pm$20 pixels. The methodology is general, however, and
it would be a straightforward process to re-derive the curve for any 
combination of processing parameters. Note that we used the same 
methodology (and the same spectrophotometric standards) to derive 
corrections to the \iue\/ Final Archive calibration \citep{massa2000}. In
addition, our calibration of the optical and NIR photometry
\citep{bstarsii}, which are used in the modeling of the stellar SEDs, was
based on these corrected \iue\/ results and, therefore, on the HST white
dwarf-based fundamental calibration. With the use of the ACS correction
derived here, all the data analyzed in this paper are based on an
internally-consistent absolute calibration.



\clearpage
\begin{deluxetable}{llrcrc}
\tablenum{1}
\tablewidth{0pc}
\tablecaption{The Target Stars and the new HST/ACS G800L Data\label{tabSTARS}}
\tablehead{
\colhead{Star} &
\colhead{Spectral} &
\colhead{V\tablenotemark{a}} &
\colhead{E(B$-$V)\tablenotemark{a}} &
\colhead{ACS} &
\colhead{Observation} \\
\colhead{} &
\colhead{Type\tablenotemark{a}} &
\colhead{} &
\colhead{} &
\colhead{Datasets} &
\colhead{Dates}}
\startdata
BD+56 517        & B1.5 V       & 10.50 & 0.51 & J9FH34041, J9FH34051 & 2005-08-02 \\
BD+45 973        & B3 V         &  8.58 & 0.80 & J9FH30041, J9FH30051 & 2005-09-01 \\
BD+44 1080       & B6 III       &  9.12 & 0.99 & J9FH29041, J9FH29051 & 2005-08-22 \\
NGC 1977 \#885   & B4 V         & 11.33 & 0.82 & J9FH15041, J9FH15051 & 2005-10-10 \\
HD 46106         & B1 V         &  7.92 & 0.43 & J9FH38041, J9FH38051 & 2005-10-19 \\
HD 292167        & O9 III:      &  9.25 & 0.72 & J9FH27041, J9FH27051 & 2005-09-03 \\
HD 68633         & B5 V         &  9.91 & 0.49 & J9FH17041, J9FH17051 & 2006-04-23 \\
HD 70614         & B6           &  9.27 & 0.68 & J9FH18041, J9FH18051 & 2006-09-21 \\
Trumpler 14 \#6  & B1 V         & 11.23 & 0.48 & J9FH52041, J9FH52051 & 2005-11-06 \\
Trumpler 14 \#27 & \nodata      & 11.30 & 0.58 & J9FH51041, J9FH51051 & 2005-08-11 \\
HD 164740        & O7.5 V(n)    & 10.30 & 0.86 & J9FH08041, J9FH08051 & 2005-08-19 \\
HD 229196        & O6 III(n)(f) &  8.52 & 1.20 & J9FH23041, J9FH23051 & 2006-03-14 \\
HD 204827        & B0 V         &  7.94 & 1.08 & J9FH11041, J9FH11051 & 2005-08-02 \\
BD+61 2365       & B0.5 V       &  9.19 & 0.78 & J9FH42041, J9FH42051 & 2005-12-22 \\
\enddata
\tablenotetext{a}{Spectral types, $V$ magnitudes, and E(B$-$V) values are
taken from Paper V; references for the spectral types are given in that
paper.} 
\end{deluxetable}


\begin{deluxetable}{lccccc}
\tabletypesize{\footnotesize}
\tablenum{2} 
\tablewidth{0pc} 
\tablecaption{NIR Extinction Curves for Target Stars
\label{tabEXTINCTION}}
\tablehead{ 
\colhead{    }                  &
\multicolumn{5}{c}{{k($\lambda$-V)}}   \\ \cline{2-6}
\colhead{Star}                  &
\colhead{$K$}                   &
\colhead{$H$}                   &
\colhead{$J$}                   &
\colhead{\it ACS92}             &
\colhead{\it ACS87}             \\
\colhead{     }                  &
\colhead{(2.18 $\mu$m)}          & 
\colhead{(1.62 $\mu$m)}          & 
\colhead{(1.23 $\mu$m)}          & 
\colhead{(9250 \AA)}             & 
\colhead{(8750 \AA)}             } 
\startdata
BD+56 517        & $-2.79\pm 0.09$ & $-2.65\pm 0.12$ & $-2.34\pm 0.08$ & $-1.95\pm 0.08$ & $-1.81\pm 0.08$  \\
BD+45 973        & $-2.47\pm 0.05$ & $-2.33\pm 0.08$ & $-2.07\pm 0.04$ & $-1.58\pm 0.05$ & $-1.46\pm 0.05$  \\
BD+44 1080       & $-2.15\pm 0.04$ & $-2.01\pm 0.06$ & $-1.80\pm 0.04$ & $-1.38\pm 0.04$ & $-1.26\pm 0.04$  \\
NGC~1977 \#885   & $-4.58\pm 0.08$ & $-4.24\pm 0.10$ & $-3.47\pm 0.06$ & $-2.43\pm 0.06$ & $-2.21\pm 0.05$  \\
HD 46106         & $-2.56\pm 0.10$ & $-2.54\pm 0.14$ & $-2.16\pm 0.09$ & $-1.64\pm 0.09$ & $-1.47\pm 0.09$  \\
HD 292167        & $-2.87\pm 0.06$ & $-2.59\pm 0.09$ & $-2.31\pm 0.05$ & $-1.78\pm 0.05$ & $-1.59\pm 0.05$  \\
HD 68633         & $-3.31\pm 0.10$ & $-3.18\pm 0.14$ & $-2.76\pm 0.08$ & $-2.17\pm 0.08$ & $-2.00\pm 0.08$ \\
HD 70614         & $-2.85\pm 0.07$ & $-2.64\pm 0.12$ & $-2.34\pm 0.06$ & $-1.84\pm 0.06$ & $-1.67\pm 0.06$ \\
Trumpler 14 \#6  & $-4.18\pm 0.12$ & $-3.85\pm 0.14$ & $-3.27\pm 0.10$ & $-2.32\pm 0.09$ & $-2.12\pm 0.09$  \\
Trumpler 14 \#27 & $-3.76\pm 0.10$ & $-3.43\pm 0.12$ & $-2.87\pm 0.08$ & $-2.05\pm 0.07$ & $-1.87\pm 0.07$  \\
HD 164740        & $-4.92\pm 0.08$ & $-4.22\pm 0.09$ & $-3.49\pm 0.06$ & $-2.48\pm 0.05$ & $-2.22\pm 0.05$  \\
HD 229196        & $-2.88\pm 0.04$ & $-2.69\pm 0.05$ & $-2.34\pm 0.03$ & $-1.74\pm 0.03$ & $-1.59\pm 0.03$ \\
HD 204827        & $-2.25\pm 0.04$ & $-2.17\pm 0.05$ & $-1.92\pm 0.03$ & $-1.51\pm 0.04$ & $-1.40\pm 0.04$  \\
BD+61 2365       &  \nodata        & $-2.68\pm 0.07$ & $-2.30\pm 0.05$ & $-1.74\pm 0.05$ & $-1.59\pm 0.05$  \\
\enddata
\end{deluxetable}

\begin{deluxetable}{lccccc}
\tabletypesize{\footnotesize}
\tablenum{2 (cont.)} 
\tablewidth{0pc} 
\tablecaption{NIR Extinction Curves for Target Stars, continued}
\tablehead{ 
\colhead{    }                  &
\multicolumn{5}{c}{{k($\lambda$-V)}}   \\ \cline{2-6}
\colhead{Star}                  &
\colhead{\it ACS82}              &
\colhead{\it ACS77}              &
\colhead{\it ACS72}              &
\colhead{\it ACS67}              &
\colhead{\it ACS62}              \\
\colhead{}                      &
\colhead{(8250 \AA)}             & 
\colhead{(7750 \AA)}             & 
\colhead{(7250 \AA)}             & 
\colhead{(6750 \AA)}             & 
\colhead{(6250 \AA)}             } 
\startdata
BD+56 517         & $-1.65\pm 0.07$ & $-1.43\pm 0.07$ & $-1.18\pm 0.07$ & $-0.92\pm 0.07$ & $-0.70\pm 0.07$ \\
BD+45 973        & $-1.31\pm 0.05$ & $-1.13\pm 0.05$ & $-0.92\pm 0.04$ & $-0.69\pm 0.04$ & $-0.44\pm 0.04$ \\
BD+44 1080        & $-1.12\pm 0.04$ & $-0.94\pm 0.04$ & $-0.77\pm 0.04$ & $-0.60\pm 0.04$ & $-0.37\pm 0.03$ \\
NGC~1977 \#885    & $-1.96\pm 0.05$ & $-1.67\pm 0.05$ & $-1.36\pm 0.05$ & $-1.02\pm 0.04$ & $-0.63\pm 0.04$ \\
HD 46106        & $-1.31\pm 0.09$ & $-1.16\pm 0.08$ & $-0.96\pm 0.08$ & $-0.76\pm 0.08$ & $-0.52\pm 0.08$ \\
HD 292167        & $-1.43\pm 0.05$ & $-1.26\pm 0.05$ & $-1.05\pm 0.05$ & $-0.82\pm 0.05$ & $-0.53\pm 0.05$ \\
HD 68633         & $-1.75\pm 0.08$ & $-1.49\pm 0.08$ & $-1.27\pm 0.07$ & $-1.06\pm 0.07$ & $-0.73\pm 0.07$ \\
HD 70614         & $-1.49\pm 0.06$ & $-1.27\pm 0.05$ & $-1.04\pm 0.05$ & $-0.82\pm 0.05$ & $-0.56\pm 0.05$ \\
Trumpler 14 \#6  & $-1.88\pm 0.08$ & $-1.62\pm 0.08$ & $-1.36\pm 0.08$ & $-1.05\pm 0.07$ & $-0.63\pm 0.07$ \\
Trumpler 14 \#27 & $-1.62\pm 0.07$ & $-1.36\pm 0.06$ & $-1.11\pm 0.06$ & $-0.85\pm 0.06$ & $-0.53\pm 0.06$ \\
HD 164740       & $-1.95\pm 0.05$ & $-1.67\pm 0.05$ & $-1.36\pm 0.04$ & $-1.04\pm 0.04$ & $-0.65\pm 0.04$ \\
HD 229196        & $-1.42\pm 0.03$ & $-1.24\pm 0.03$ & $-1.04\pm 0.03$ & $-0.82\pm 0.03$ & $-0.56\pm 0.03$ \\
HD 204827        & $-1.26\pm 0.04$ & $-1.10\pm 0.03$ & $-0.92\pm 0.03$ & $-0.71\pm 0.03$ & $-0.43\pm 0.03$ \\
BD+61 2365       & $-1.41\pm 0.05$ & $-1.21\pm 0.05$ & $-1.00\pm 0.05$ & $-0.77\pm 0.04$ & $-0.50\pm 0.04$ \\
\enddata
\end{deluxetable}


\begin{deluxetable}{lcccccccccc}
\tabletypesize{\scriptsize}
\tablenum{3}
\tablewidth{0pc}
\tablecaption{Results for Power-Law Fits\label{tabRESULTS}}
\tablehead{
\colhead{Star} &
\multicolumn{4}{c}{$\beta$--Fixed\tablenotemark{a}} &
\colhead{} &
\multicolumn{4}{c}{$\beta$--Variable\tablenotemark{b}} &
\colhead{F\tablenotemark{c}} \\ \cline{2-5} \cline{7-10}
\colhead{} &
\colhead{$k_{IR}$} &
\colhead{$\beta$} &
\colhead{$R(V)$} &
\colhead{$\chi^2$} &
\colhead{} &
\colhead{$k_{IR}$} &
\colhead{$\beta$} &
\colhead{$R(V)$} &
\colhead{$\chi^2$} &
\colhead{ }}
\startdata
BD+56 517        & $ 0.98^{+ 0.06}_{-0.04}$ & $ 1.84$ & $ 3.03^{+ 0.07}_{-0.07}$ & $ 0.17$ & & $ 0.87^{+ 0.15}_{-0.10}$ & $ 2.10^{+ 0.18}_{-0.29}$ & $ 2.95^{+ 0.13}_{-0.07}$ & $ 0.16$& $ 1.1$ \\
BD+45 973        & $ 1.00^{+ 0.04}_{-0.02}$ & $ 1.84$ & $ 2.73^{+ 0.04}_{-0.03}$ & $ 0.07$ & & $ 0.93^{+ 0.06}_{-0.06}$ & $ 2.00^{+ 0.15}_{-0.11}$ & $ 2.68^{+ 0.06}_{-0.06}$ & $ 0.03$& $ 8.8$ \\
BD+44 1080       & $ 0.89^{+ 0.03}_{-0.02}$ & $ 1.84$ & $ 2.39^{+ 0.03}_{-0.03}$ & $ 0.29$ & & $ 0.76^{+ 0.05}_{-0.04}$ & $ 2.21^{+ 0.15}_{-0.12}$ & $ 2.28^{+ 0.04}_{-0.04}$ & $ 0.02$& $ 63.6$ \\
NGC 1977 \#885   & $ 2.13^{+ 0.04}_{-0.04}$ & $ 1.84$ & $ 4.99^{+ 0.05}_{-0.07}$ & $ 2.40$ & & $ 2.97^{+ 0.18}_{-0.09}$ & $ 1.26^{+ 0.04}_{-0.08}$ & $ 5.74^{+ 0.17}_{-0.10}$ & $ 0.64$& $ 14.9$ \\
HD 46106         & $ 1.09^{+ 0.07}_{-0.05}$ & $ 1.84$ & $ 2.88^{+ 0.09}_{-0.07}$ & $ 0.23$ & & $ 1.00^{+ 0.15}_{-0.10}$ & $ 2.03^{+ 0.30}_{-0.20}$ & $ 2.81^{+ 0.17}_{-0.10}$ & $ 0.26$& $ 0.4$ \\
HD 292167        & $ 1.18^{+ 0.03}_{-0.04}$ & $ 1.84$ & $ 3.12^{+ 0.04}_{-0.05}$ & $ 0.24$ & & $ 1.28^{+ 0.10}_{-0.08}$ & $ 1.66^{+ 0.10}_{-0.13}$ & $ 3.21^{+ 0.09}_{-0.08}$ & $ 0.22$& $ 1.4$ \\
HD 68633         & $ 1.35^{+ 0.06}_{-0.04}$ & $ 1.84$ & $ 3.69^{+ 0.08}_{-0.07}$ & $ 0.24$ & & $ 1.14^{+ 0.10}_{-0.09}$ & $ 2.24^{+ 0.19}_{-0.16}$ & $ 3.52^{+ 0.10}_{-0.10}$ & $ 0.11$& $ 6.5$ \\
HD 70614         & $ 1.15^{+ 0.04}_{-0.03}$ & $ 1.84$ & $ 3.13^{+ 0.06}_{-0.05}$ & $ 0.13$ & & $ 1.06^{+ 0.09}_{-0.07}$ & $ 2.02^{+ 0.14}_{-0.21}$ & $ 3.06^{+ 0.09}_{-0.07}$ & $ 0.12$& $ 1.8$ \\
Trumpler 14 \#6  & $ 1.90^{+ 0.08}_{-0.05}$ & $ 1.84$ & $ 4.58^{+ 0.12}_{-0.07}$ & $ 0.34$ & & $ 2.34^{+ 0.20}_{-0.17}$ & $ 1.44^{+ 0.10}_{-0.11}$ & $ 4.98^{+ 0.20}_{-0.12}$ & $ 0.12$& $ 10.7$ \\
Trumpler 14 \#27 & $ 1.75^{+ 0.06}_{-0.04}$ & $ 1.84$ & $ 4.12^{+ 0.08}_{-0.06}$ & $ 0.35$ & & $ 2.16^{+ 0.21}_{-0.11}$ & $ 1.45^{+ 0.09}_{-0.13}$ & $ 4.48^{+ 0.20}_{-0.11}$ & $ 0.03$& $ 53.7$ \\
HD 164740        & $ 2.26^{+ 0.04}_{-0.04}$ & $ 1.84$ & $ 5.17^{+ 0.06}_{-0.06}$ & $ 5.51$ & & $ 4.27^{+ 0.39}_{-0.24}$ & $ 0.89^{+ 0.04}_{-0.06}$ & $ 7.04^{+ 0.38}_{-0.25}$ & $ 0.10$& $283.1$ \\
HD 229196        & $ 1.23^{+ 0.02}_{-0.02}$ & $ 1.84$ & $ 3.17^{+ 0.03}_{-0.03}$ & $ 0.42$ & & $ 1.29^{+ 0.04}_{-0.04}$ & $ 1.72^{+ 0.07}_{-0.06}$ & $ 3.23^{+ 0.04}_{-0.05}$ & $ 0.43$& $ 1.0$ \\
HD 204827        & $ 0.86^{+ 0.03}_{-0.02}$ & $ 1.84$ & $ 2.49^{+ 0.04}_{-0.02}$ & $ 0.46$ & & $ 0.73^{+ 0.04}_{-0.04}$ & $ 2.24^{+ 0.13}_{-0.10}$ & $ 2.38^{+ 0.04}_{-0.03}$ & $ 0.11$& $ 16.5$ \\
BD+61 2365       & $ 1.21^{+ 0.05}_{-0.03}$ & $ 1.84$ & $ 3.14^{+ 0.05}_{-0.05}$ & $ 0.09$ & & $ 1.38^{+ 0.34}_{-0.17}$ & $ 1.61^{+ 0.28}_{-0.34}$ & $ 3.30^{+ 0.32}_{-0.16}$ & $ 0.05$& $ 3.7$ \\
\enddata
\tablenotetext{a}{``Universal" value of $\beta = 1.84$ assumed in these
fits. Reduced-$\chi^2$ values are shown, corresponding to 5 degrees of
freedom. Fits for the sight line towards BD+61 2365, for which no $K$-band
data are available, have 4 degrees of freedom.}
\tablenotetext{b}{Power-law exponent $\beta$ taken as a free parameter in
these fits. Reduced-$\chi^2$ values are shown, corresponding to 4 degrees
of freedom. Fits for the sight line towards BD+61 2365 have 3 degrees of
freedom.}
\tablenotetext{c}{The F-statistic, which tests the need to include $\beta$
as a free parameter in the power law fits \citep{bevington1969}.}
\end{deluxetable}


\begin{deluxetable}{lcccc}
\tabletypesize{\scriptsize}
\tablenum{4} 
\tablewidth{0pc} 
\tablecaption{Results for Fits Using Equation~(\ref{eqnPEI2})\tablenotemark{a}\label{tabPEI}}
\tablehead{ 
\colhead{Star}                           &  
\colhead{$\alpha$}                       &
\colhead{$R(V)$}                         &
\colhead{$\chi^2$}                       &
\colhead{$\chi^2$}                       \\ 
\colhead{}                               &  
\colhead{}                               &
\colhead{}                               &
\colhead{}                               &
\colhead{($\beta$--Variable)}            }
\startdata
BD+56 517        & $ 2.89^{+ 0.12}_{-0.15}$ & $ 2.86^{+ 0.07}_{-0.07}$ & $ 0.20$ & $ 0.14$ \\
BD+45 973        & $ 2.49^{+ 0.08}_{-0.10}$ & $ 2.64^{+ 0.06}_{-0.04}$ & $ 0.13$ & $ 0.23$ \\
BD+44 1080       & $ 2.47^{+ 0.06}_{-0.10}$ & $ 2.30^{+ 0.05}_{-0.03}$ & $ 0.30$ & $ 0.69$ \\
NGC 1977 \#885   & $ 1.73^{+ 0.05}_{-0.06}$ & $ 5.54^{+ 0.12}_{-0.11}$ & $ 0.38$ & $ 0.87$ \\
HD 46106         & $ 2.45^{+ 0.20}_{-0.25}$ & $ 2.77^{+ 0.12}_{-0.08}$ & $ 0.17$ & $ 0.33$ \\
HD 292167        & $ 2.40^{+ 0.10}_{-0.09}$ & $ 3.04^{+ 0.06}_{-0.07}$ & $ 0.21$ & $ 0.22$ \\
HD 68633         & $ 2.54^{+ 0.09}_{-0.15}$ & $ 3.51^{+ 0.11}_{-0.08}$ & $ 0.38$ & $ 0.60$ \\
HD 70614         & $ 2.48^{+ 0.09}_{-0.10}$ & $ 3.02^{+ 0.07}_{-0.06}$ & $ 0.22$ & $ 0.48$ \\
Trumpler 14 \#6  & $ 1.94^{+ 0.08}_{-0.11}$ & $ 4.80^{+ 0.21}_{-0.16}$ & $ 0.10$ & $ 0.24$ \\
Trumpler 14 \#27 & $ 1.81^{+ 0.08}_{-0.10}$ & $ 4.45^{+ 0.19}_{-0.11}$ & $ 0.18$ & $ 0.43$ \\
HD 164740        & $ 1.61^{+ 0.04}_{-0.05}$ & $ 6.00^{+ 0.15}_{-0.12}$ & $ 0.15$ & $ 0.20$ \\
HD 229196        & $ 2.34^{+ 0.05}_{-0.05}$ & $ 3.09^{+ 0.04}_{-0.04}$ & $ 0.75$ & $ 1.27$ \\
HD 204827        & $ 2.75^{+ 0.07}_{-0.09}$ & $ 2.35^{+ 0.04}_{-0.02}$ & $ 0.13$ & $ 0.24$ \\
BD+61 2365       & $ 2.28^{+ 0.11}_{-0.12}$ & $ 3.11^{+ 0.11}_{-0.09}$ & $ 0.08$ & $ 0.19$ \\
\enddata
\tablenotetext{a}{The final column is $\chi^2$ for fits using the $\beta$-Variable model 
applied to the same wavelength range used in the fits employing equation~(\ref{eqnPEI2}).}
\end{deluxetable}


\clearpage

\begin{figure}[ht]
\figurenum{3}
\plotone{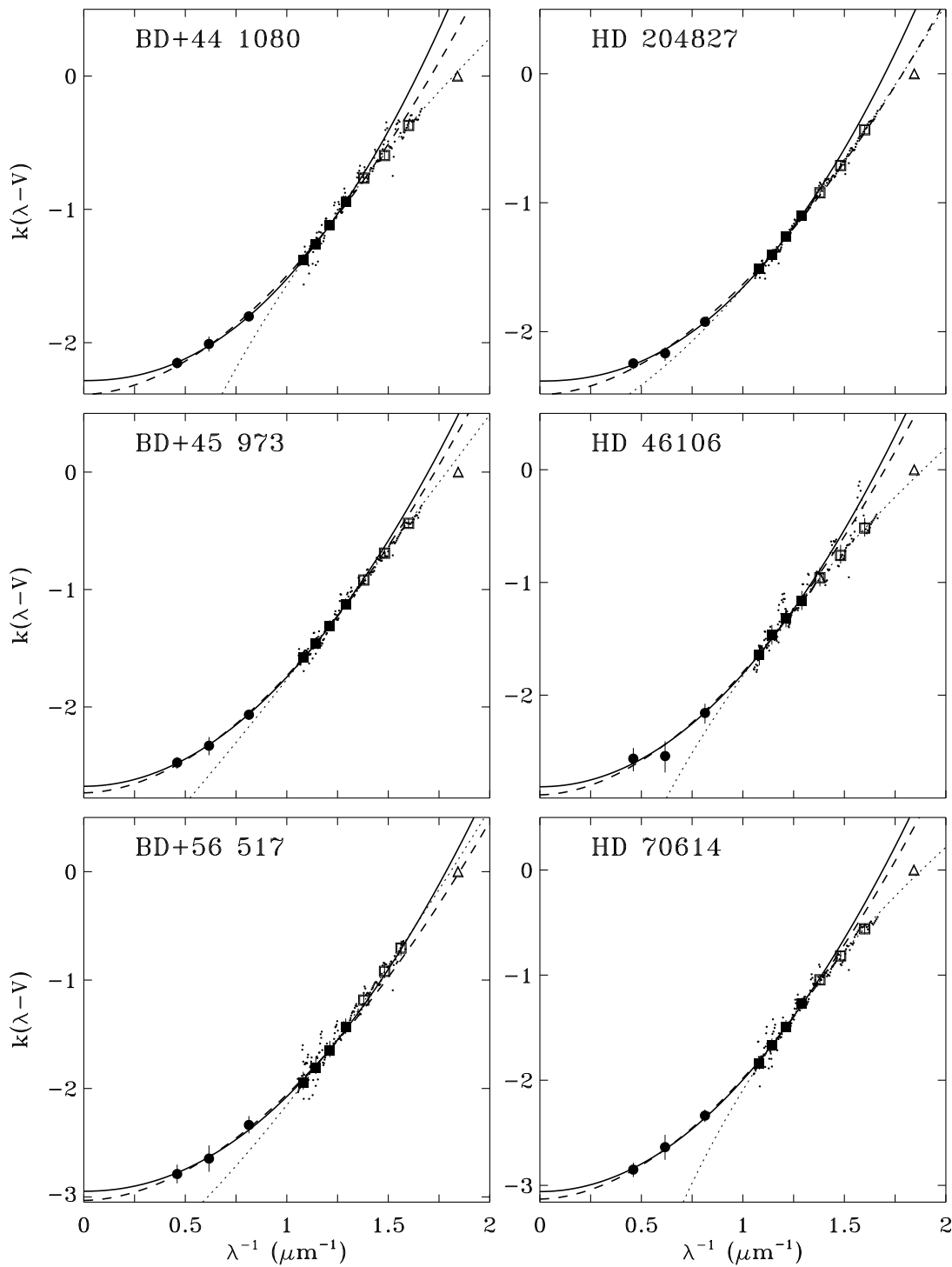}
\caption{NIR extinction curves for six of the 14 program sight lines. The 
different symbols show the \klam\/ for the \tmass\/ \jhk\/ photometry 
(large circles), ACS spectrophotometry (open and filled squares), and 
Johnson $V$ magnitudes (triangles), as listed in Table~2. 
The ACS data were binned into a set of seven magnitudes before computing
the extinction. The unbinned ASC data are shown as small circles (see 
\S~\ref{sec_CURVES}). The dashed curves are the fits to the \jhk\/ and ACS 
data using Equation~(\ref{eqnIR2}) with $\beta \equiv 1.84$ (``$\beta$--Fixed'' 
results in Table~3). The solid curves are fits using
Equation~(\ref{eqnIR2}) with $\beta$ as a free parameter (``$\beta$--Variable'' 
results in Table~3). For both sets of fits, only
ACS data longward of 7500 \AA\/  ($\lambda^{-1} < 1.33$ \invmic) were included (filled squares).  Shorter
wavelength data (open squares) are not consistent with a single power-law 
representation.  The dotted curves are $\beta$--Variable fits to the ACS 
data only.  The curves are organized in order of increasing $R(V)$.
\label{fig_CURVES}}
\end{figure}

\begin{figure}[ht]
\figurenum{3 (cont.)}
\plotone{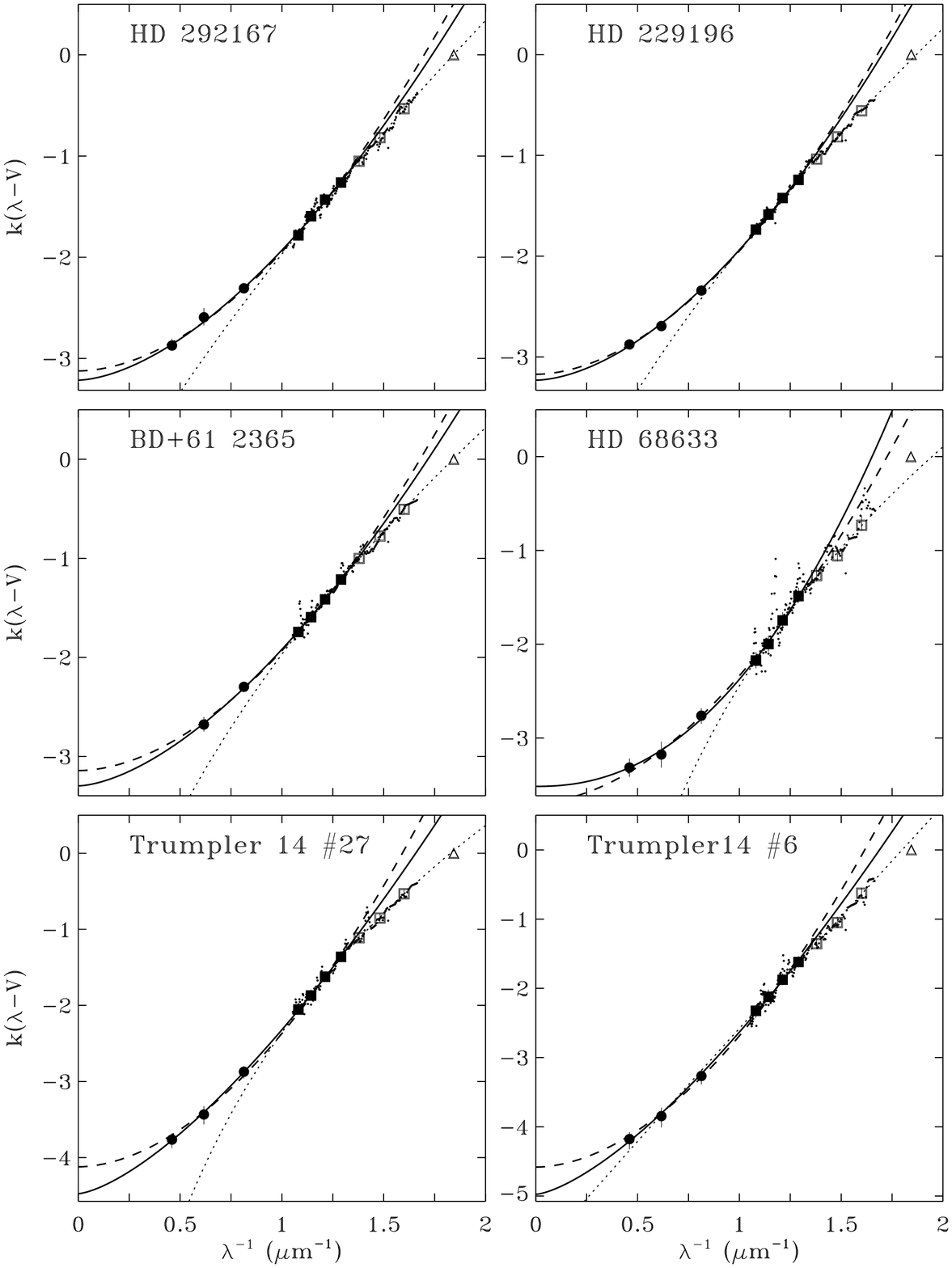}
\caption{Continuation of Figure \ref{fig_CURVES} for six more sight 
lines.}
\end{figure}

\begin{figure}[ht]
\figurenum{3 (cont.)}
\plotone{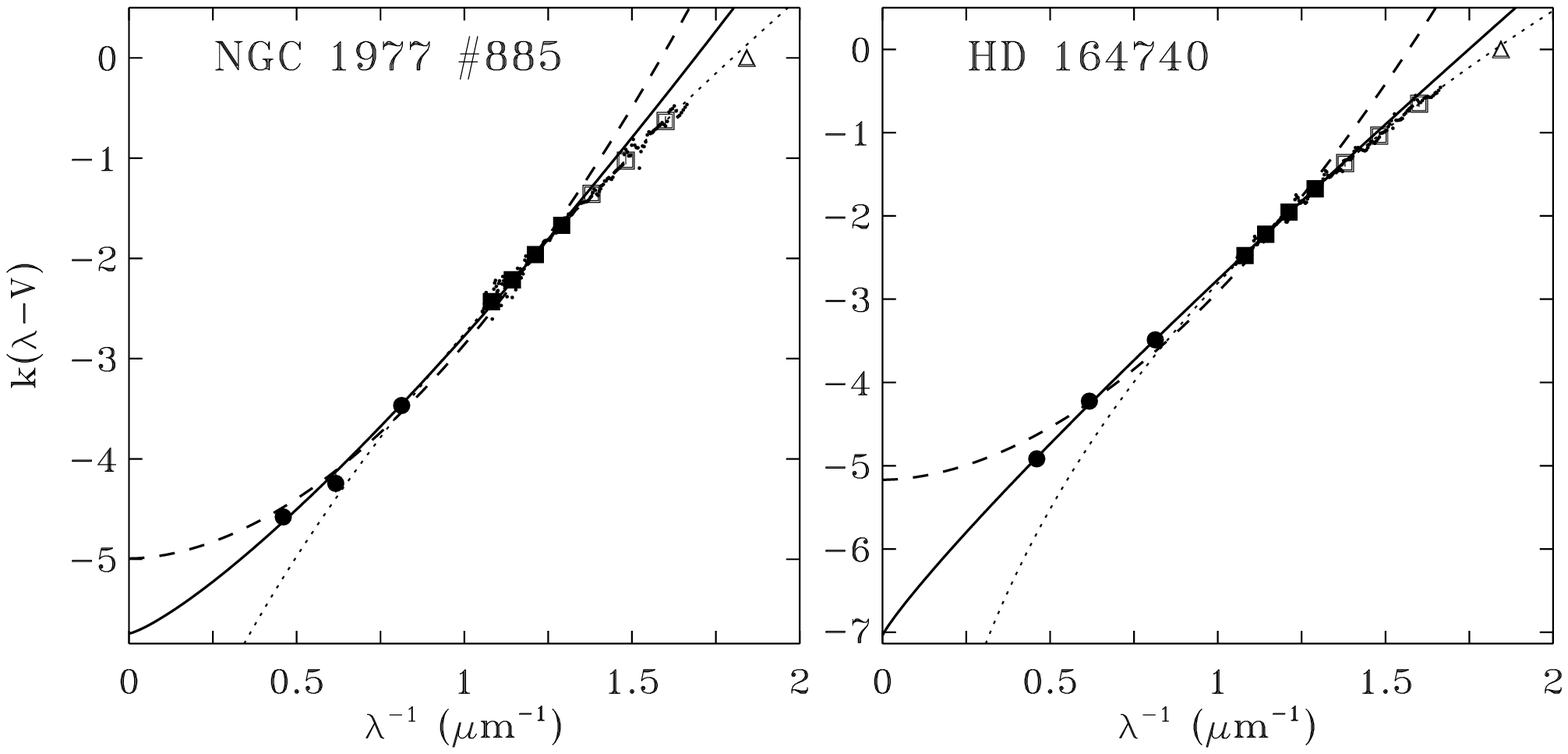}
\caption{Continuation of Figure \ref{fig_CURVES} for the two highest 
$R(V)$ sight lines.}
\end{figure}

\begin{figure}[ht]
\figurenum{4}
\plotone{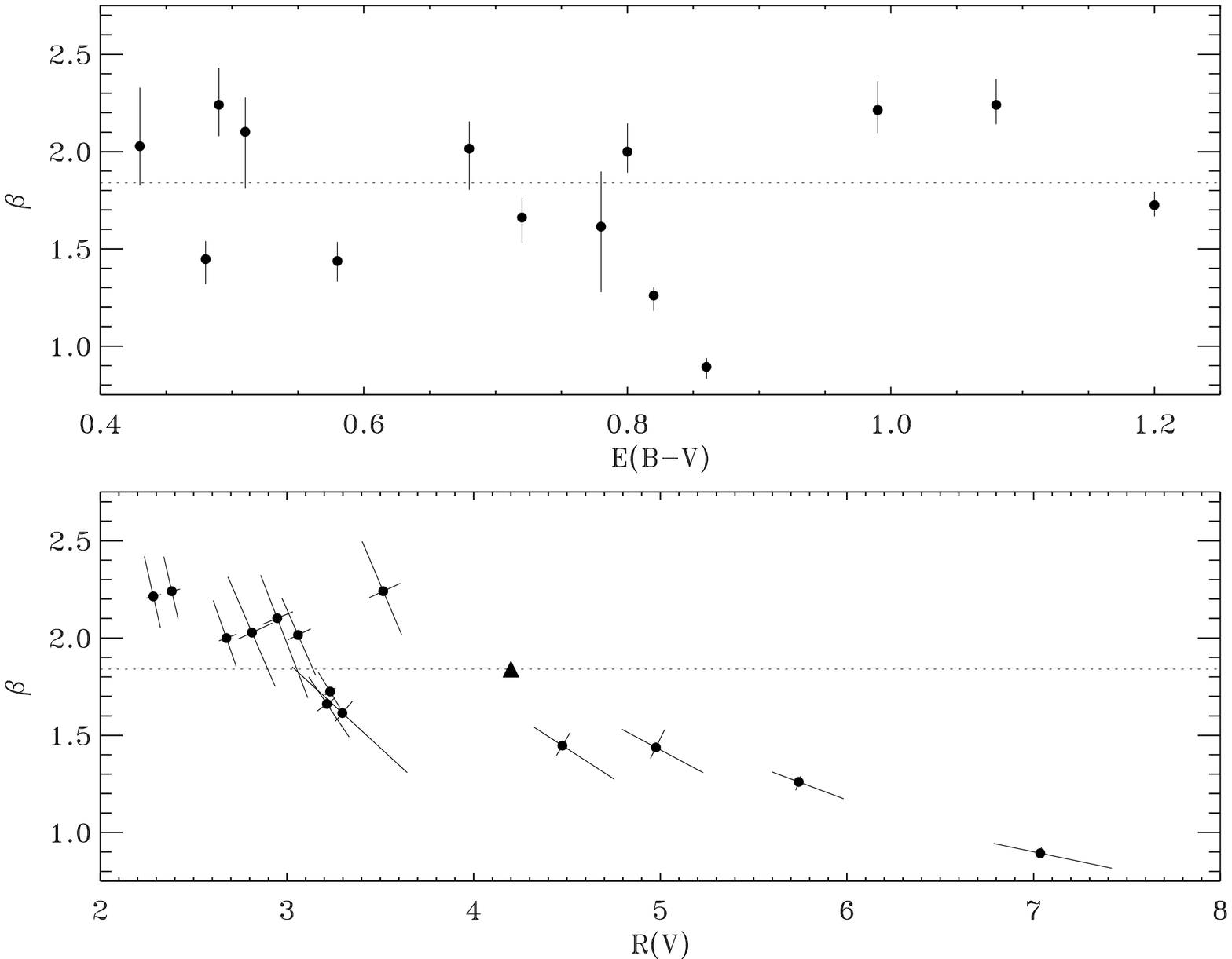}
\caption{The power-law exponent $\beta$\ versus \ebv\/ (top panel) and
$R(V)$ (bottom panel) for the ``$\beta$--Variable'' fits from Table
3. The dotted horizontal lines show the value $\beta =
1.84$. The filled triangle in the lower panel represents the sight line to
the star $\rho$ Oph from \citealt{martin1990}. The 1-$\sigma$ error bars
are based on fits to the Monte Carlo simulations of the data and the noise
model.
\label{fig_BETA}}
\end{figure}

\begin{figure}[ht]
\figurenum{5}
\plotone{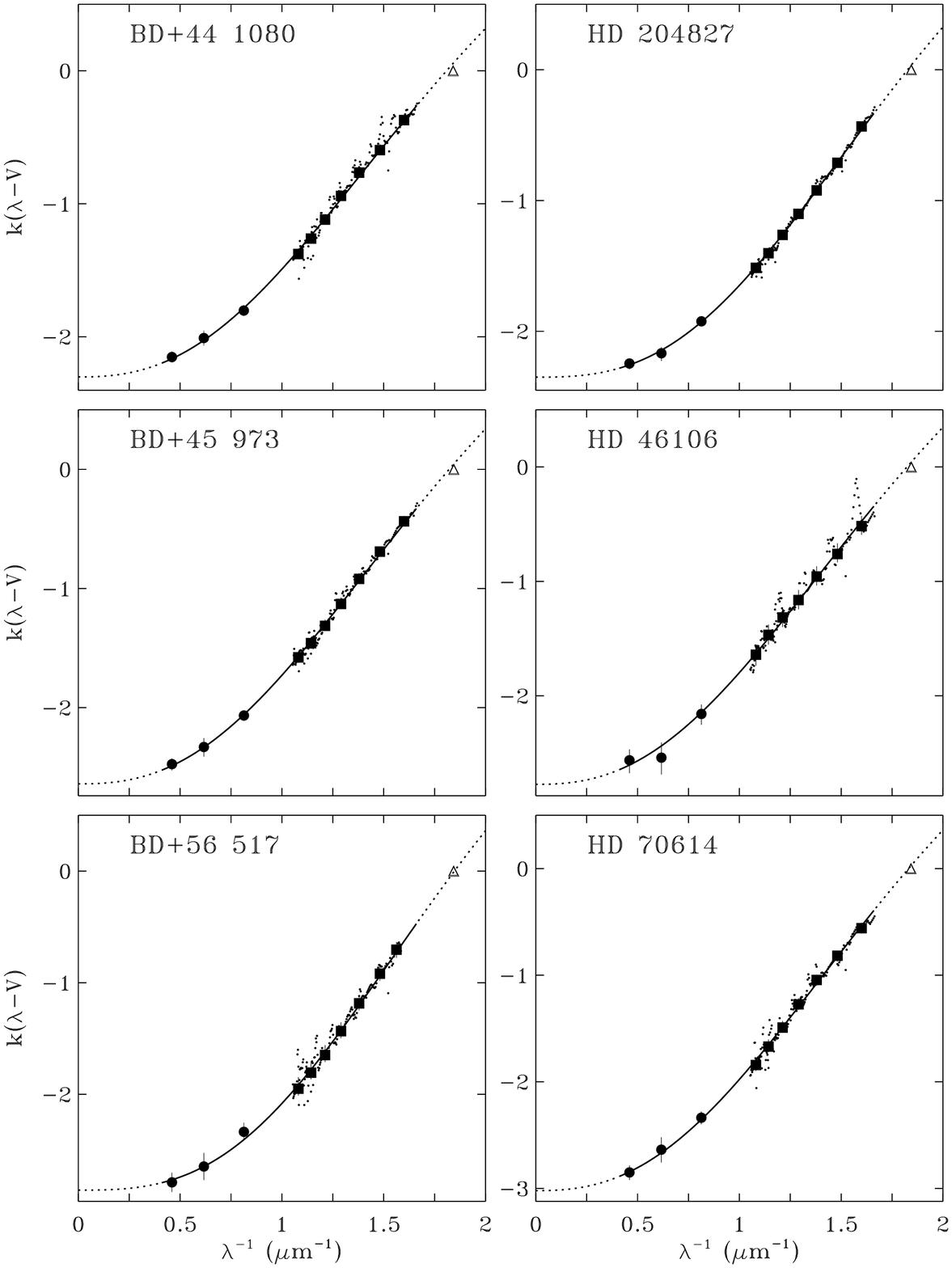}
\caption{Observed NIR extinction curves compared to fits determined from 
Equation~(\ref{eqnPEI2}). The data and symbols are the same as in 
Figure~\ref{fig_CURVES}.  All the ACS data were 
included in the fits (thus all the large squares are shown as filled). The
dotted portions of the curves indicate extrapolations of the fitting 
function beyond the range of the data.  The sight lines are organized in 
order of increasing $R(V)$.
\label{fig_CURVES_PEI_a}}
\end{figure}

\begin{figure}[ht]
\figurenum{5 (cont.)}
\plotone{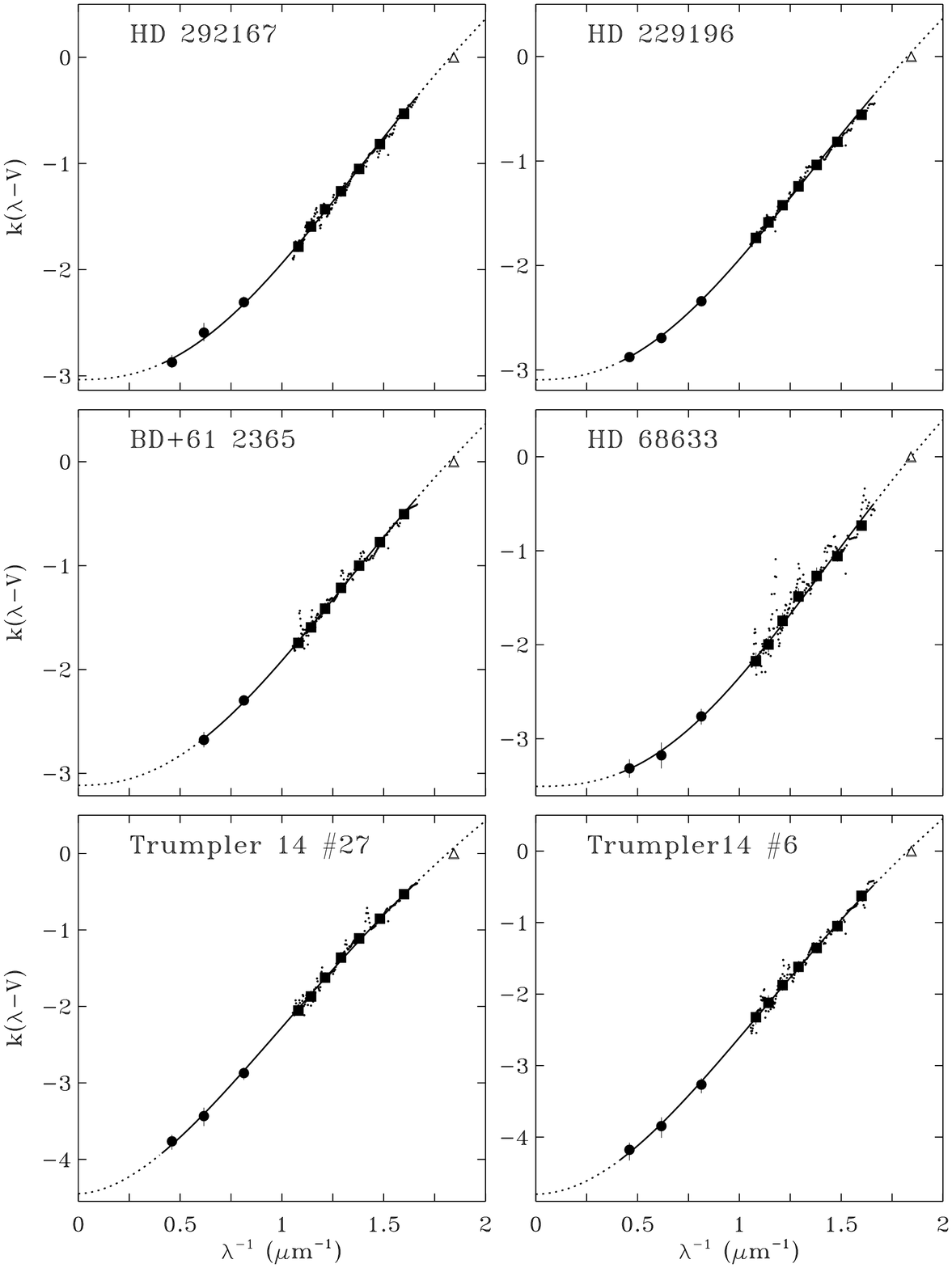}
\caption{Continuation of Figure \ref{fig_CURVES_PEI_a} for six more sight lines.
\label{fig_CURVES_PEI_b}}
\end{figure}

\begin{figure}[ht]
\figurenum{5 (cont.)}
\plotone{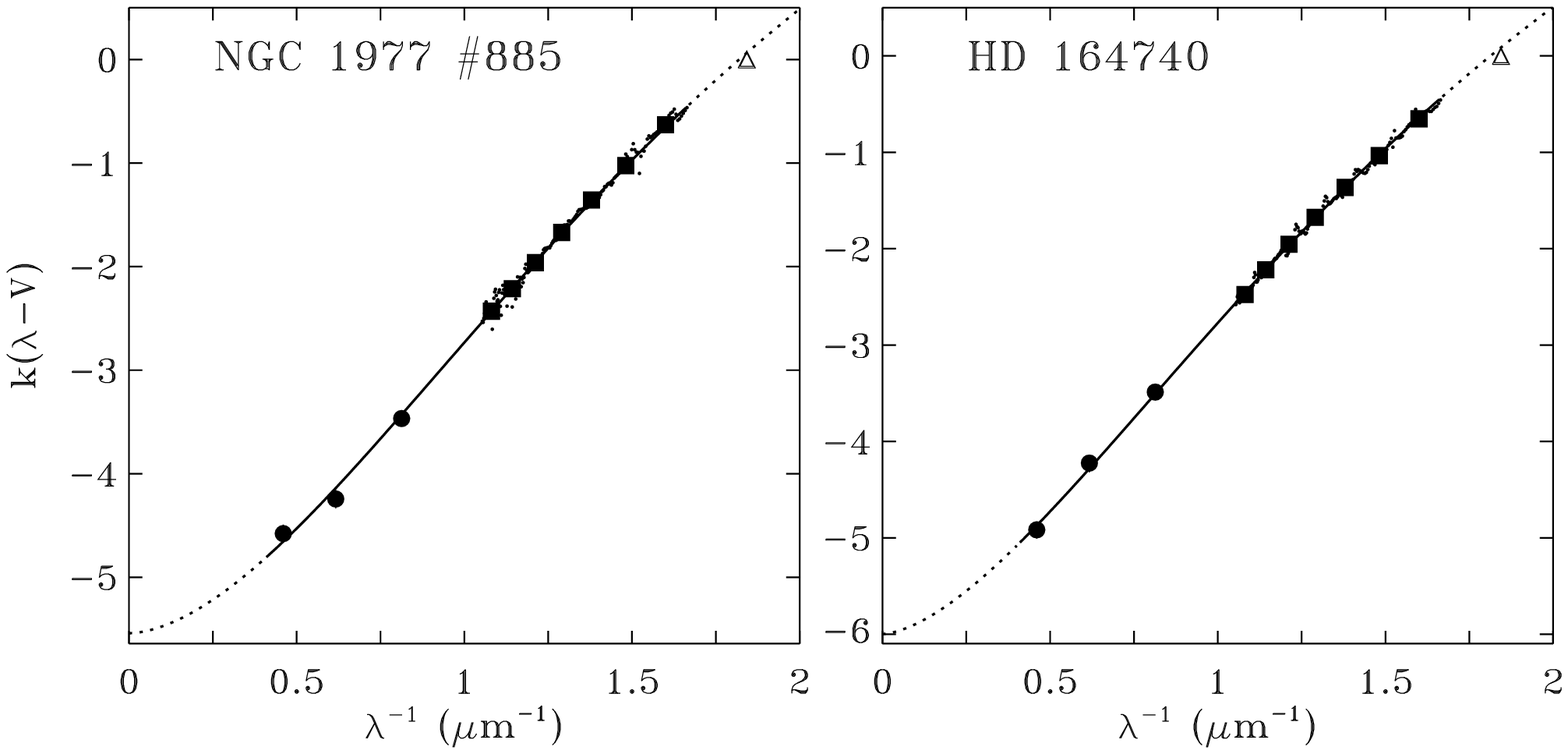}
\caption{Continuation of Figure \ref{fig_CURVES_PEI_a} for the two highest $R(V)$ sight lines.
\label{fig_CURVES_PEI_c}}
\end{figure}

\begin{figure}[ht]
\figurenum{6}
\plotone{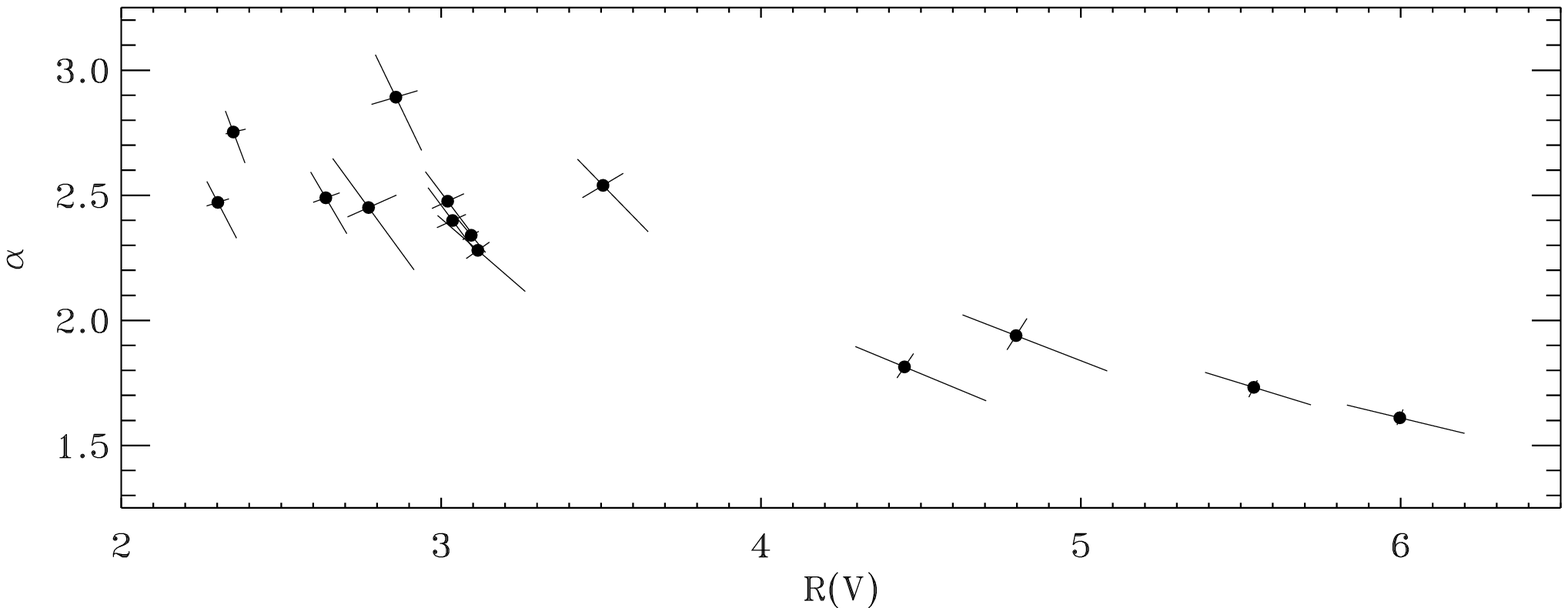}
\caption{Plot of the $\alpha$ and $R(V)$ from Table 4.  The 
1-$\sigma$ error bars are based on fits to the Monte Carlo simulations 
of the data and the noise model.
\label{fig_BETA_PEI}}
\end{figure}

\begin{figure}[ht]
\figurenum{7}
\plotone{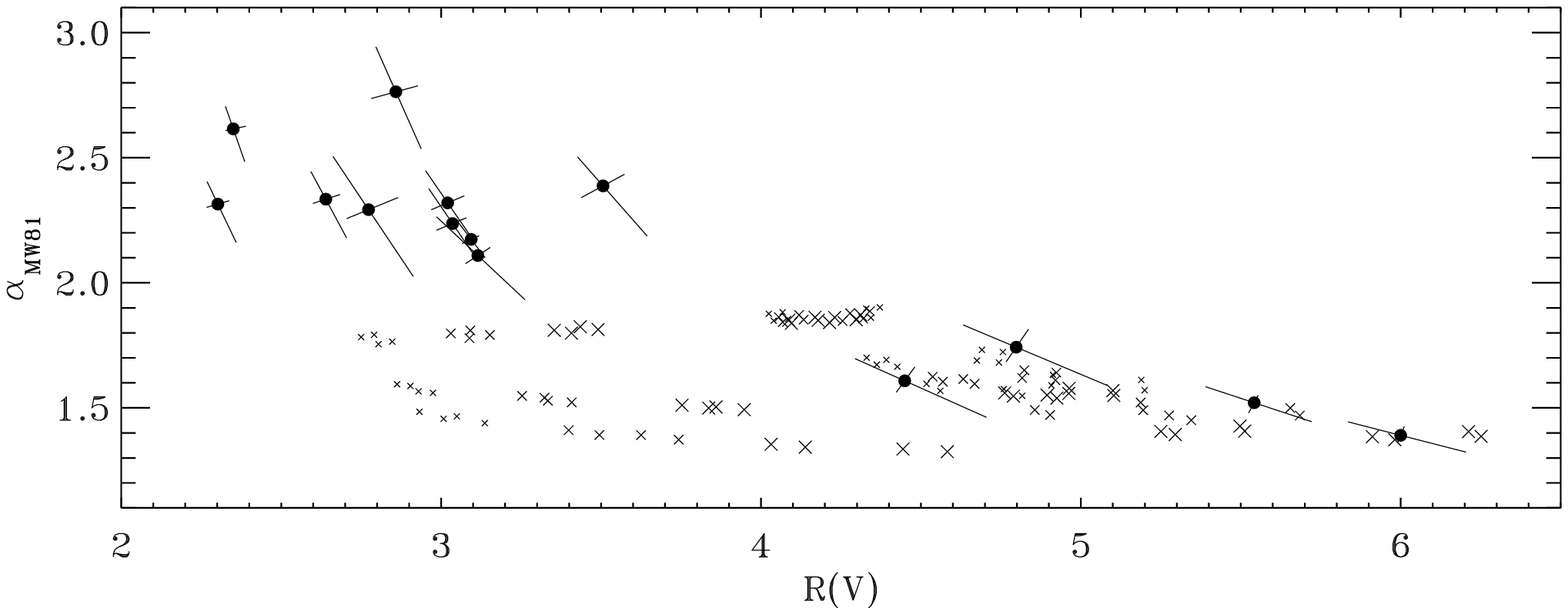}
\caption{Comparison of extinction curve parameters derived from observed 
and theoretical extinction curves.  $\alpha_{MW81}$ is the exponent that 
results from a simple power-law fit (see eq.~[\ref{eqnIR2}]) to the IR 
extinction at $\lambda^{-1}$ = 0.00, 0.29, and 1.11 \invmic.  It provides a 
measure of the shape of the curves for $\lambda > 9000$ \AA.  The crosses 
show $\alpha_{MW81}$ and $R(V)$ calculated from the \citet{mathis1981} 
theoretical extinction curves for a variety of size distributions and 
abundance ratios (see text). The small, medium, and large crosses 
correspond to (C/Si)$_{dust}$ = 3.25, 6.5, and 13.0, respectively. The 
filled circles show the same quantities derived from the best-fitting NIR 
curves shown in Figure~\ref{fig_CURVES_PEI_a}. 
This figure shows that some combinations of model dust abundance and size 
distribution are not favored by nature.  Most notable, are the crosses in 
the lower left of the figure, which does not contain observed points.  The 
models in this region correspond to size distributions with small upper 
cutoffs for silicates ($a_+$ = 0.25 \mic) and large upper cutoffs for 
graphite ($a_+$ = 0.40 or 0.50 \mic).  The \citet{mathis1981} models are 
not useful for addressing the shapes of the smaller $R(V)$ sight lines.
\label{fig_BETA_PEI_2}}
\end{figure}

\begin{figure}[ht]
\figurenum{8}
\plotone{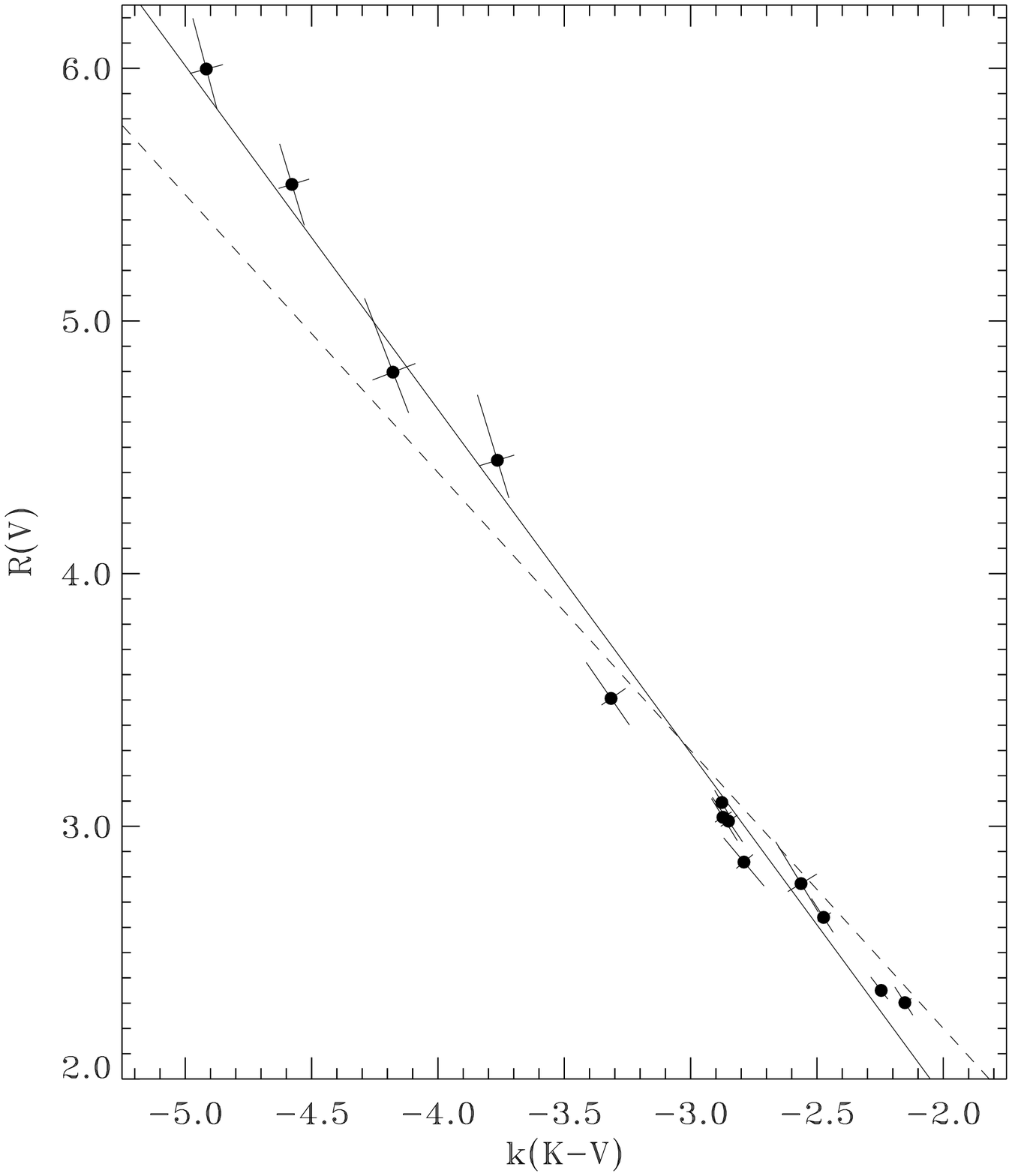}
\caption{Plot of the observed $k (K-V) \equiv E(K-V)/E(B-V)$ versus the $R(V)$ derived 
from fits 
to Eq.~(\ref{eqnPEI2}).  The 1-$\sigma$ error bars illustrate the 
correlated errors in the two quantities.  The dashed line gives the 
often-used relation $R(V) = -1.1 \frac{E(K-V)}{E(B-V)}$, which is based on van de Hulst's 
theoretical extinction curve No.\ 15 \citep[see, e.g.,][]{Johnson1968}.
An unweighted least squares linear fit to the data is shown as the solid 
line, $R(V) = -1.36 \frac{E(K-V)}{E(B-V)} - 0.79$.
\label{fig_KLAMK}}
\end{figure}

\begin{figure}[ht]
\figurenum{A1}
\plotone{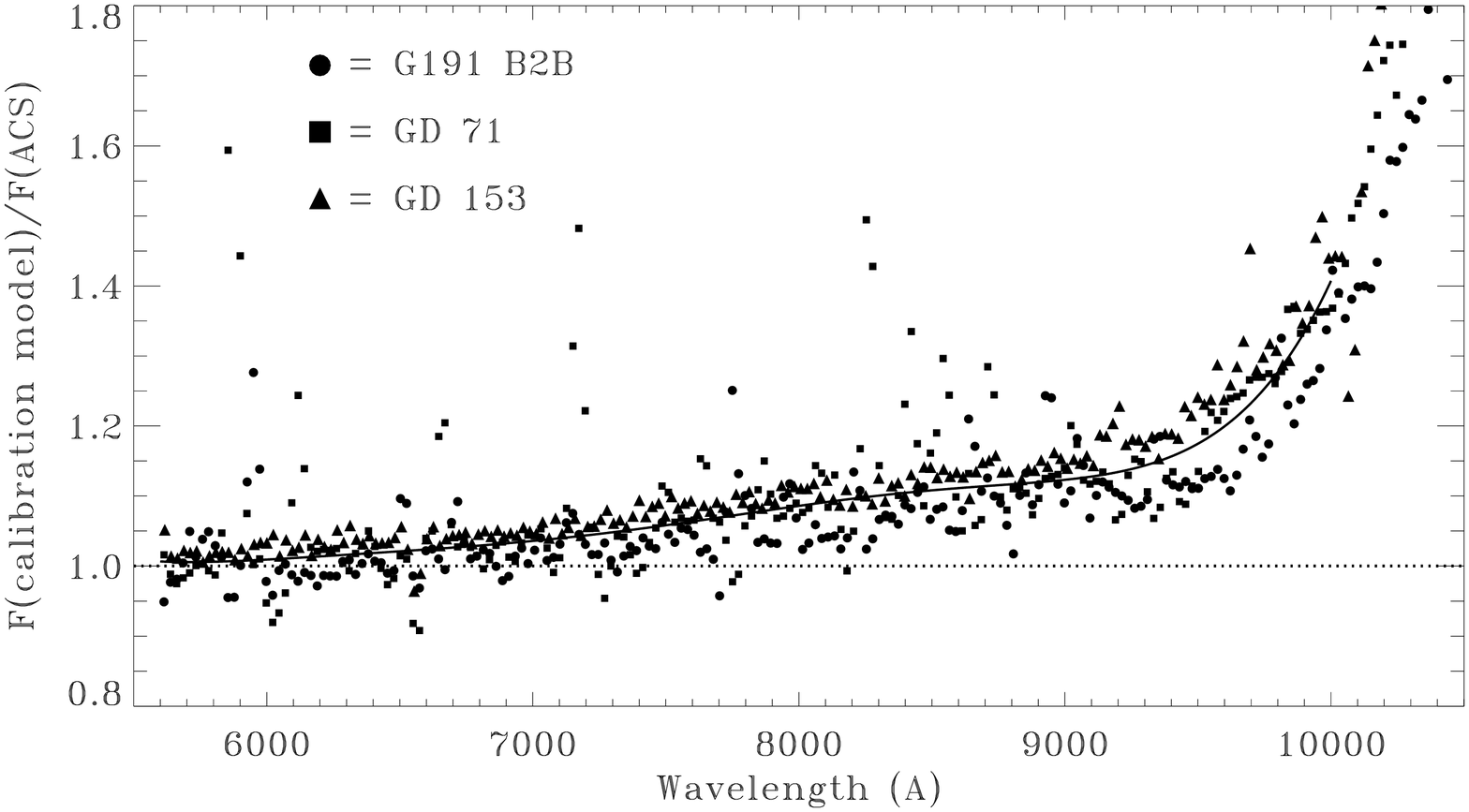}
\caption{Correction to the ACS/HRC/G800L calibration derived from the three 
fundamental calibration stars G191 B2B, GD~71, and GD~153.
\label{fig_CALSPEC}}
\end{figure}

\begin{figure}[ht]
\figurenum{A2}
\plotone{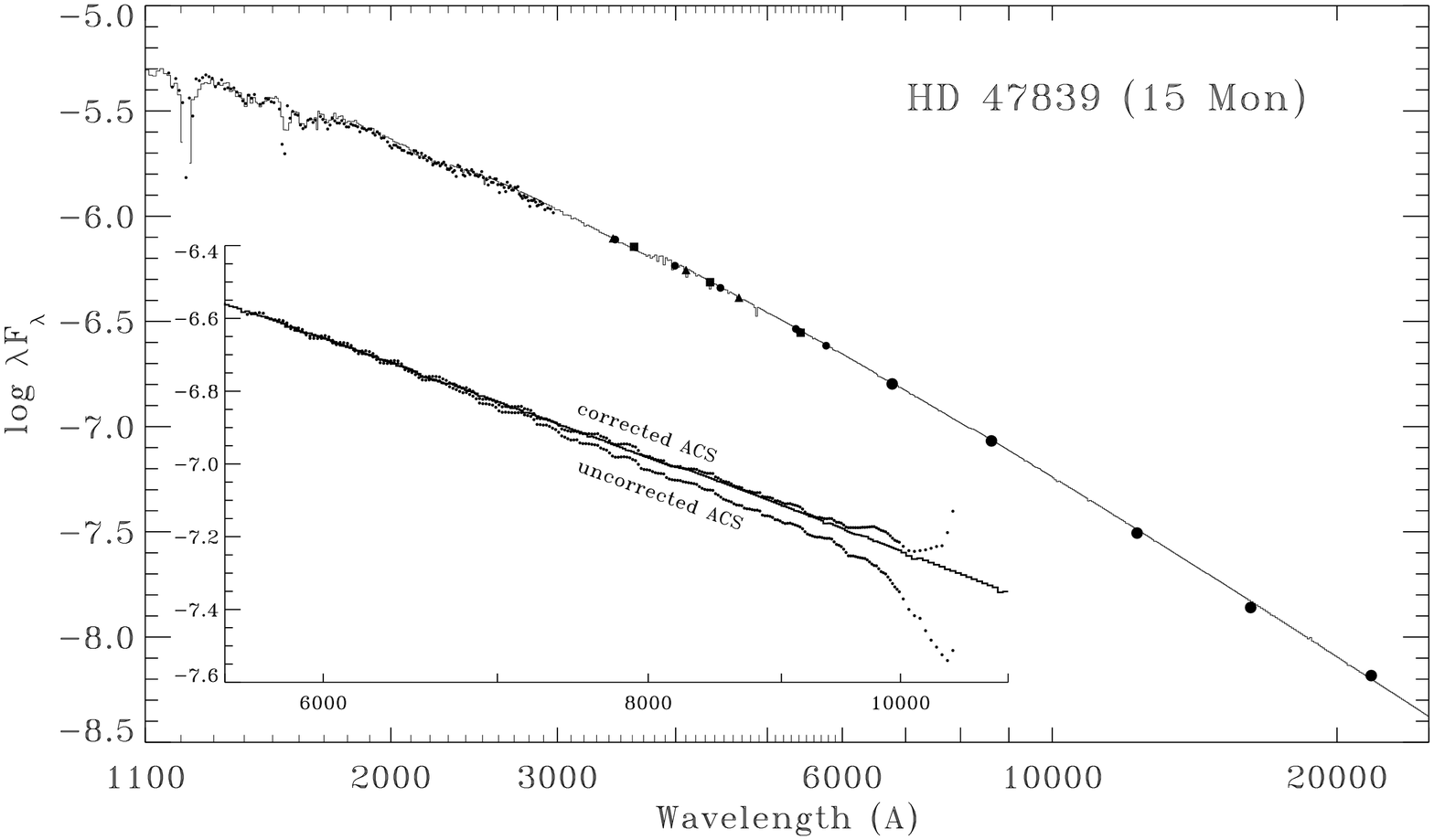}
\caption{Verification of the ACS/HRC/G800L calibration correction from
Figure \ref{fig_CALSPEC}. The main figure shows the observed SED of
HD~47839, including low-resolution \iue\/ spectrophotometry (small circles)
in the UV ($\lambda < 3000$ \AA), \ubv\/ (large circles), Str\"{o}mgren
{\it uvby} (triangles), and Geneva {\it UB$_1$B$_2$V$_1$G} (diamonds)
photometry in the optical (3000 \AA\/ $ < \lambda < 6000$ \AA), and {\it
RIJHK} photometry in the NIR ($\lambda > 6000$ \AA). The smooth curve is a
37,000 K TLUSTY model atmosphere \citep{lanz2003} fit to the data from
\citep[e.g.,][]{FMIV, bstarsii}. The inset illustrates the effect of the
ACS/HRC/G800L correction curve shown in Figure \ref{fig_CALSPEC}. It shows
the best fit model from the main figure (continuous curve) with the
corrected and uncorrected G800L data (points) overplotted. For $\lambda
\leq 9500$~\AA\/ the re-calibrated data mimic the shape of the model SED,
but lie $\sim 1$\% above it, consistent with the scatter among the
spectrophotometric standards. For $\lambda \geq 9500$~\AA, the fluxes are
clearly much less reliable \label{fig_15MON}}
\end{figure}


\begin{thebibliography}{32}
\expandafter\ifx\csname natexlab\endcsname\relax\def\natexlab#1{#1}\fi

\bibitem[{{Bevington}(1969)}]{bevington1969}
{Bevington}, P.~R. 1969, {Data reduction and error analysis for the physical
  sciences} (New York: McGraw-Hill, 1969)

\bibitem[{{Bohlin}(1996)}]{bohlin1996}
{Bohlin}, R.~C. 1996, \aj, 111, 1743

\bibitem[{{Cardelli} {et~al.}(1989){Cardelli}, {Clayton}, \& {Mathis}}]{CCM}
{Cardelli}, J.~A., {Clayton}, G.~C., \& {Mathis}, J.~S. 1989, \apj, 345, 245

\bibitem[{{Clayton} {et~al.}(2003){Clayton}, {Gordon}, {Salama}, {Allamandola},
  {Martin}, {Snow}, {Whittet}, {Witt}, \& {Wolff}}]{clayton2003b}
{Clayton}, G.~C., {Gordon}, K.~D., {Salama}, F., {Allamandola}, L.~J.,
  {Martin}, P.~G., {Snow}, T.~P., {Whittet}, D.~C.~B., {Witt}, A.~N., \&
  {Wolff}, M.~J. 2003, \apj, 592, 947

\bibitem[{{Fitzpatrick} \& {Massa}(1986)}]{FMI}
{Fitzpatrick}, E.~L. \& {Massa}, D. 1986, \apj, 307, 286

\bibitem[{{Fitzpatrick} \& {Massa}(1988)}]{FMII}
---. 1988, \apj, 328, 734

\bibitem[{{Fitzpatrick} \& {Massa}(1990)}]{FMIII}
---. 1990, \apjs, 72, 163

\bibitem[{{Fitzpatrick} \& {Massa}(2005{\natexlab{a}})}]{FMIV}
---. 2005{\natexlab{a}}, \aj, 130, 1127

\bibitem[{{Fitzpatrick} \& {Massa}(2005{\natexlab{b}})}]{bstarsii}
---. 2005{\natexlab{b}}, \aj, 129, 1642

\bibitem[{{Fitzpatrick} \& {Massa}(2007)}]{FMV}
---. 2007, \apj, 663, 320

\bibitem[{{Froebrich} {et~al.}(2007){Froebrich}, {Murphy}, {Smith}, {Walsh}, \&
  {Del Burgo}}]{froebrich2007}
{Froebrich}, D., {Murphy}, G.~C., {Smith}, M.~D., {Walsh}, J., \& {Del Burgo},
  C. 2007, \mnras, 378, 1447

\bibitem[{{Gordon}(2009)}]{Gordon2009}
{Gordon}, K. 2009, in preparation

\bibitem[{{Gosling} {et~al.}(2009){Gosling}, {Bandyopadhyay}, \&
  {Blundell}}]{gosling2009}
{Gosling}, A.~J., {Bandyopadhyay}, R.~M., \& {Blundell}, K.~M. 2009, ArXiv
  e-prints

\bibitem[{{Hecht} {et~al.}(1982){Hecht}, {Helfer}, {Wolf}, {Pipher}, \&
  {Donn}}]{hecht1982}
{Hecht}, J., {Helfer}, H.~L., {Wolf}, J., {Pipher}, J.~L., \& {Donn}, B. 1982,
  \apjl, 263, L39

\bibitem[{{Johnson}(1968)}]{Johnson1968}
{Johnson}, H.~L. 1968, {Interstellar Extinction} (Nebulae and interstellar
  matter.~Edited by Barbara M.~Middlehurst; Lawrence H.~Aller.~Library of
  Congress Catalog Card Number 66-13879.~Published by the University of Chicago
  Press, Chicago, ILL USA, 1968, p.167), Chapter 5

\bibitem[{{Kurucz}(1991)}]{kurucz1991}
{Kurucz}, R.~L. 1991, in NATO ASIC Proc. 341: Stellar Atmospheres - Beyond
  Classical Models, 441

\bibitem[{{Lanz} \& {Hubeny}(2003)}]{lanz2003}
{Lanz}, T. \& {Hubeny}, I. 2003, \apjs, 146, 417

\bibitem[{{Larsen} {et~al.}(2006){Larsen}, {K{\"u}mmel}, \&
  {Walsh}}]{larsen2006}
{Larsen}, S.~S., {K{\"u}mmel}, M., \& {Walsh}, J.~R. 2006, in The 2005 HST
  Calibration Workshop: Hubble After the Transition to Two-Gyro Mode, ed. A.~M.
  {Koekemoer}, P.~{Goudfrooij}, \& L.~L. {Dressel}, 103

\bibitem[{{Larson} \& {Whittet}(2005)}]{larson2005}
{Larson}, K.~A. \& {Whittet}, D.~C.~B. 2005, \apj, 623, 897

\bibitem[{{Markwardt}(2009)}]{Markwardt2009}
{Markwardt}, C.~B. 2009, ArXiv e-prints

\bibitem[{{Martin} \& {Whittet}(1990)}]{martin1990}
{Martin}, P.~G. \& {Whittet}, D.~C.~B. 1990, \apj, 357, 113

\bibitem[{{Massa} \& {Fitzpatrick}(2000)}]{massa2000}
{Massa}, D. \& {Fitzpatrick}, E.~L. 2000, \apjs, 126, 517

\bibitem[{{Mathis} {et~al.}(1977){Mathis}, {Rumpl}, \& {Nordsieck}}]{MRN}
{Mathis}, J.~S., {Rumpl}, W., \& {Nordsieck}, K.~H. 1977, \apj, 217, 425

\bibitem[{{Mathis} \& {Wallenhorst}(1981)}]{mathis1981}
{Mathis}, J.~S. \& {Wallenhorst}, S.~G. 1981, \apj, 244, 483

\bibitem[{{Nichols} \& {Linsky}(1996)}]{nichols1996}
{Nichols}, J.~S. \& {Linsky}, J.~L. 1996, \aj, 111, 517

\bibitem[{{Nishiyama} {et~al.}(2006){Nishiyama}, {Nagata}, {Kusakabe},
  {Matsunaga}, {Naoi}, {Kato}, {Nagashima}, {Sugitani}, {Tamura}, {Tanab{\'e}},
  \& {Sato}}]{nishiyama2006}
{Nishiyama}, S., {Nagata}, T., {Kusakabe}, N., {Matsunaga}, N., {Naoi}, T.,
  {Kato}, D., {Nagashima}, C., {Sugitani}, K., {Tamura}, M., {Tanab{\'e}}, T.,
  \& {Sato}, S. 2006, \apj, 638, 839

\bibitem[{{Pei}(1992)}]{Pei1992}
{Pei}, Y.~C. 1992, \apj, 395, 130

\bibitem[{{Rieke} \& {Lebofsky}(1985)}]{rieke1985}
{Rieke}, G.~H. \& {Lebofsky}, M.~J. 1985, \apj, 288, 618

\bibitem[{{Spitzer}(1978)}]{Spitzer1978}
{Spitzer}, L. 1978, {Physical processes in the interstellar medium} (New York
  Wiley-Interscience, 1978.)

\bibitem[{{Walborn}(1973)}]{Walborn1973}
{Walborn}, N.~R. 1973, \aj, 78, 1067

\bibitem[{{Whittet}(2003)}]{whittet2003}
{Whittet}, D.~C.~B. 2003, {Dust in the galactic environment} ({Bristol:
  Institute of Physics (IOP) Publishing})

\bibitem[{{Zubko} {et~al.}(2004){Zubko}, {Dwek}, \& {Arendt}}]{zubko2004}
{Zubko}, V., {Dwek}, E., \& {Arendt}, R.~G. 2004, \apjs, 152, 211

\end{thebibliography}
\end{document}